\documentclass[aps,prd,twocolumn,superscriptaddress,nofootinbib,10pt]{revtex4-1}

\usepackage[utf8]{inputenc}
\usepackage{graphicx,amsmath,amsfonts,amssymb,aas_macros,slashed}
\usepackage{bbold}
\usepackage{datetime}
\usepackage{numprint}
\usepackage{enumitem}
\usepackage{graphicx,color,amsmath}
\usepackage{hyperref}

\usepackage{rotating}
\allowdisplaybreaks

\newcommand{\MeV}{\ensuremath{\mathrm{MeV}}}
\newcommand{\GeV}{\ensuremath{\mathrm{GeV}}}
\newcommand{\TeV}{\ensuremath{\mathrm{TeV}}}

\usepackage{pgfplots}
\usepackage{bm}
\usepackage{mathtools}

\makeatletter
\newcommand{\thickhline}{%
    \noalign {\ifnum 0=`}\fi \hrule height 1pt
    \futurelet \reserved@a \@xhline
}
\newcolumntype{"}{@{\hskip\tabcolsep\vrule width 1pt\hskip\tabcolsep}}
\makeatother

\begin{document}

\title{Scalar Dark Matter Candidates -- Revisited}

\author{C\'eline B\oe hm}
\email{celine.boehm@sydney.edu.au}
\affiliation{School of Physics, Physics Road, The University of Sydney, NSW 2006 Camperdown, Sydney, Australia}

\author{Xiaoyong Chu}
\email{xiaoyong.chu@oeaw.ac.at}
\affiliation{Institute of High Energy Physics, Austrian Academy of Sciences, Nikolsdorfergasse 18, 1050 Vienna, Austria}
\author{Jui-Lin Kuo}
\email{jui-lin.kuo@oeaw.ac.at}
\affiliation{Institute of High Energy Physics, Austrian Academy of Sciences, Nikolsdorfergasse 18, 1050 Vienna, Austria}
\author{Josef Pradler}
\email{josef.pradler@oeaw.ac.at}
\affiliation{Institute of High Energy Physics, Austrian Academy of Sciences, Nikolsdorfergasse 18, 1050 Vienna, Austria}

\begin{abstract}
We revisit the possibility of light scalar dark matter, in the MeV to GeV mass bracket and coupled to electrons through fermion or vector mediators, in light of significant experimental and observational advances that probe new physics below the GeV-scale. We establish new limits from electron colliders and fixed-target beams, and derive the strength of loop-induced processes that are probed by precision physics, among other laboratory probes. In addition, we compute the cooling bound from SN1987A, consider self-scattering, structure formation, and cosmological constraints as well as the limits from dark matter-electron scattering in direct detection experiments. We then show that the combination of constraints largely excludes the possibility that the galactic annihilation of these particles may explain the long-standing INTEGRAL excess of 511~keV photons as observed in the galactic bulge. As caveat to these conclusions we identify the resonant annihilation regime where the vector mediator goes nearly on-shell. 
\end{abstract}

\maketitle

\section{Introduction}

The history of scalar dark matter (DM) in the MeV-GeV mass bracket is a long one. It has its roots when main-stream literature was primarily focusing on  electroweak-scale new (supersymmetric) physics. High-energy colliders had long explored the GeV-scale, and naive cosmological considerations suggested that thermal DM needed to have a mass of several GeV for the least~\cite{Hut:1977zn,Lee:1977ua}.
However, light-scalar DM~\cite{Boehm:2002yz,Boehm:2003hm} turned out to be a perfect possibility---and still is. For example, it can couple to the Standard Model (SM) fermions either by a Yukawa-type interaction of some heavy fermions $F$ or by a new gauge interaction mediated by a new vector particle $Z'$~\cite{Boehm:2003hm}. 

Cosmologically, such DM retained its right of existence by achieving a sufficient annihilation cross section through an equally light $Z'$ (which has an even longer history~\cite{Fayet:1980ad,Fayet:1980rr}), or by a possible near independence of DM mass in the annihilation cross section when $F$ is involved~\cite{Boehm:2003hm}. Experimentally, a light $Z'$ was viable, because the neutral current phenomenology remained largely unaffected, either from a suppression with center-of-mass energy arising from momentum-resolved diagrams in processes at high-energy, or, at low-energy, by ensuring that the effective strength of the interaction is smaller than the weak interactions of the SM. In turn, a model with $F$-mediation was even simpler to retain as the mass-scale of these particles can be in the~TeV. 

This, at the time seeming ``niche physics'' quickly gained momentum. When the SPI spectrometer on board of the INTEGRAL satellite confirmed a strong flux of 511~keV photons at the level of almost $ 10^{-3}\,/{\rm cm}^2/{\rm s}$~\cite{Jean:2003ci, Knodlseder:2003sv}  (see also \cite{Prantzos:2010wi,Siegert:2015knp}) coming from the galactic bulge, a MeV-scale DM origin was suggested on the basis of its spatial morphology and its general compatibility with the relic density requirement~\cite{Boehm:2003bt} while at the same time obeying soft gamma-ray constraints~\cite{Boehm:2002yz,Beacom:2004pe,Rasera:2005sa}. Concretely, the signal, especially its high bulge-to-disk ratio, is unexpected from known astrophysics~\cite{Prantzos:2010wi,Panther:2017flc} and calls for a new production mechanism of low-energy positrons. This can be achieved through DM annihilation into $e^+ e^-$ pairs.

The second piece of early impetus for such models was their connection to low-energy precision physics, in particular to the electron and muon anomalous magnetic moments $(g-2)_{e,\mu}$. There is a curious coincidence between the DM-viable parameters of the models, in part suggested by~\eqref{Eq:INTEGRAL_values}, and the lepton-mass dependent shifts to $(g-2)_{e(\mu)}$ on the order of $10^{-11}$ ($10^{-9}$)---essentially at the level of their observed magnitudes~\cite{Boehm:2003bt}.%
\footnote{The more general phenomenology of a light $Z'$, including coupling to quarks, was originally considered in~\cite{Fayet:1980ad,Fayet:1980rr,Fayet:1986rh}.}
This is a nice example of  how  laboratory probes of SM quantities inform us on the astrophysical and cosmological viability of new physics and vice versa~\cite{Boehm:2004gt}.

A lot of general phenomenological progress has been made since the introduction of the aforementioned scalar DM models. Regarding the INTEGRAL interpretation, the DM mass is now strongly constrained, e.g.~from annihilation in flight, and the DM-mass now generally needs to be below tens of~MeV, and can even be fully excluded in specific models and/or under certain assumptions of the state of the early Universe; for more details see, e.g.,~\cite{Boehm:2002yz,Beacom:2004pe,Boehm:2006df, Wilkinson:2016gsy} and references therein. 
In turn, the measurement~\cite{Bennett:2006fi} of the anomalous magnetic moment of the muon now stands in (3--4)$\sigma$ tension with the SM predicted one~\cite{Jegerlehner:2009ry}.
In fact, the imminent experimental update for $(g-2)_{\mu}$ is much in the limelight today, especially after its connection with GeV-scale dark sector physics became more broadly appreciated following~\cite{Pospelov:2008zw}.
Finally, a beyond the Standard Model (BSM) sector that contains light dark states has found further motivations, such as the astrophysical core/cusp problem~\cite{Moore:1994yx, Flores:1994gz, Spergel:1999mh}, various galactic cosmic ray excesses~\cite{Goodenough:2009gk,Pospelov:2008jd,ArkaniHamed:2008qn,Hooper:2010mq}, among others. These motivations paired with the to-date absence of new physics at the electroweak  (EW) scale have acted as a great innovation driver for devising laboratory and observational tests for sub-GeV dark sector physics, see, e.g.~\cite{Essig:2013lka, Alexander:2016aln,Battaglieri:2017aum} and references therein.

\begin{figure*}[t]
    \centering
    \includegraphics[width=0.85\linewidth]{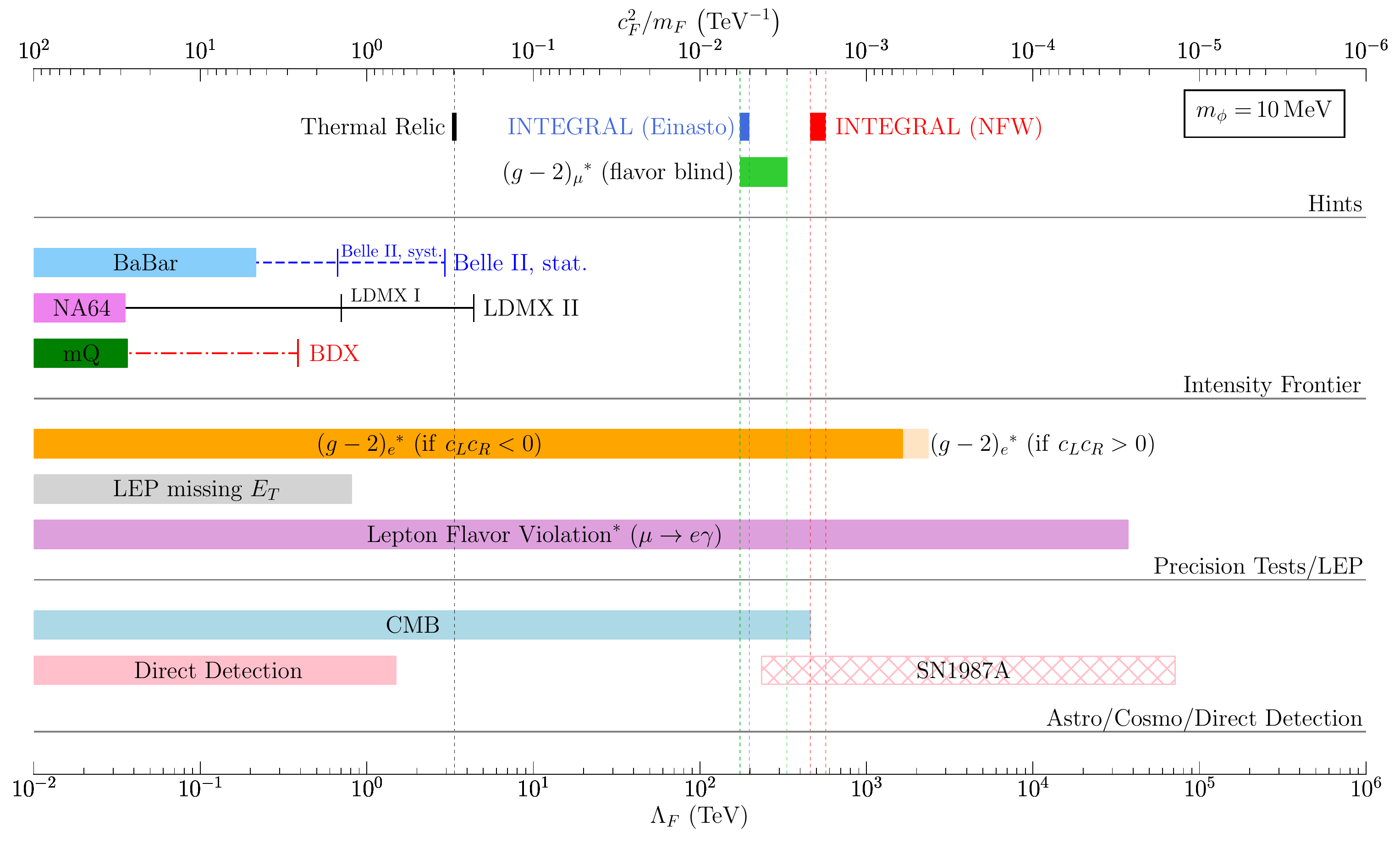}
    \caption{Summary of constraints obtained in this paper for the fermion-mediated model as a function of the effective UV-scale $\Lambda_F = (c_F^2/m_F)^{-1}$ for a fixed DM mass of $m_\phi = 10\,\MeV$  in heavy mediator limit; $c_F = \sqrt{|c_L^e c_R^e|}$. A star indicates, that the bound only applies under certain conditions. The top section ``{Hints}'' shows the regions of interest for the explanation of the  INTEGRAL signal, for the $(g-2)_\mu$ anomaly assuming flavor-blind couplings and same $F$ masses between the first two generations, and the point for achieving the correct relic density through DM freeze-out. The next section ``{Intensity Frontier}'' shows constraints (projections) from searches for missing momentum in $e^+e^- $ collisions at BaBar (Belle II), for missing energy in the $e^-$ fixed target experiment NA64 (LDMX), and for direct $\phi$-$e^-$ scattering of $\phi$ produced in the $e^-$ fixed target experiment mQ (BDX). The section ``{Precision Tests/LEP}'' shows the  conservative constraint from the loop-induced contribution to $(g-2)_e$ for either sign of the product of couplings as labeled~\cite{Parker:2018vye,Morel:2020dww}, the limit on missing energy searches at LEP and, in the case of a single generation of $F$ and assuming flavor-blind couplings, the limit from the lepton flavor violating $\mu\to e \gamma$ transition. The final section ``{Astro/Cosmo/Direct Detection}'' is devoted to CMB limits on energy injection, to direct detection limits from $\phi$-electron scattering and from anomalous energy loss in SN1987A. Weaker limits such as from the invisible width of the $Z$, from structure formation, from DM self-scattering, from the running of $\alpha$, from the left-right asymmetry in polarized electron-electron scattering are not shown (see main text instead).   The INTEGRAL interpretation is excluded.}
    \label{Fig:summary_F}
\end{figure*}

\begin{figure*}[t]
    \centering
    \includegraphics[width=0.85\linewidth]{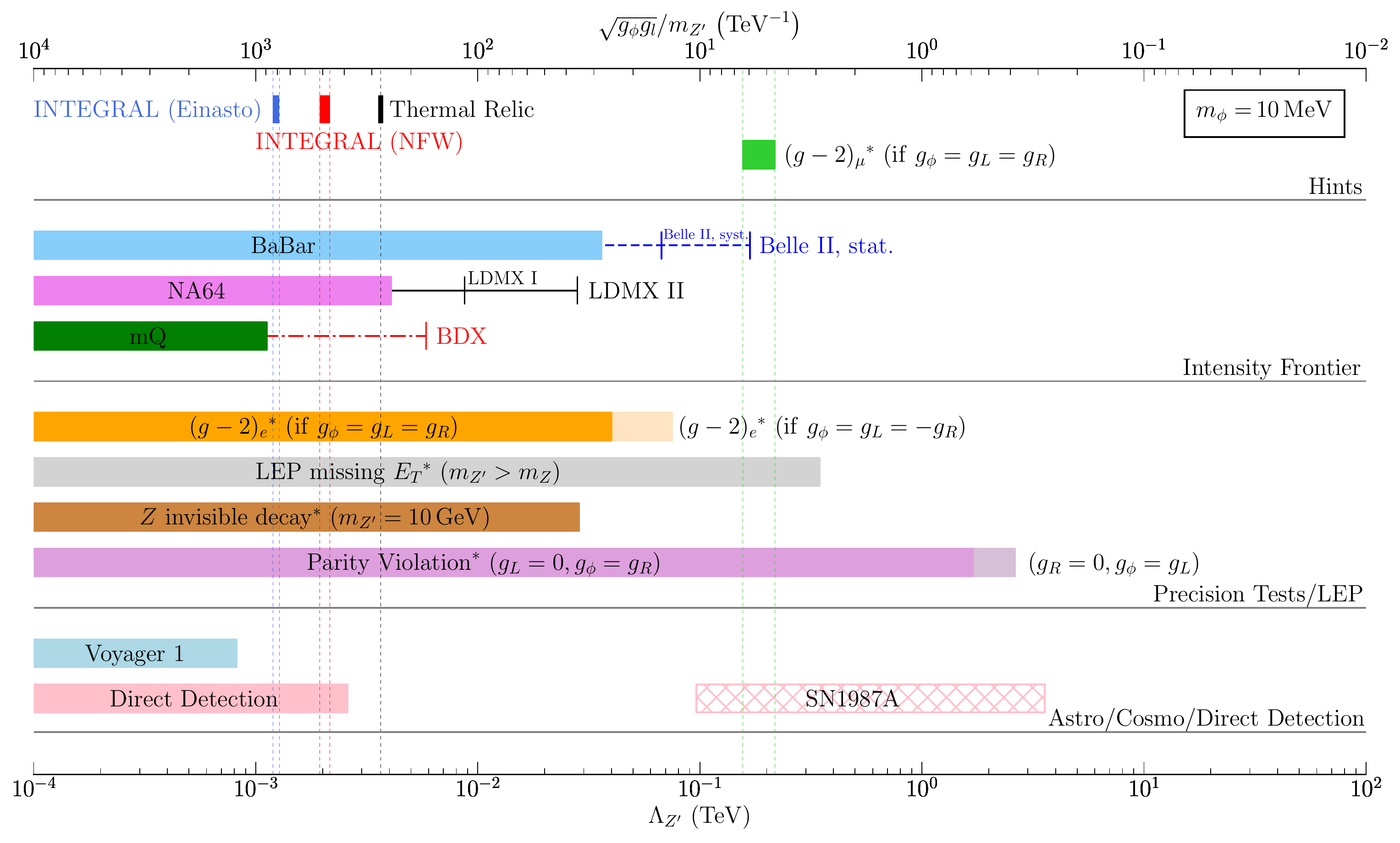}
    \caption{Summary of constraints obtained in this paper for the vector-mediated model as a function of the effective scale $\Lambda_{Z'} = (\sqrt{g_\phi g_l}/m_{Z'})^{-1}$ for a fixed DM mass of $m_\phi = 10\,\MeV$ in heavy mediator limit, similar to Fig.~\ref{Fig:summary_F}. A star indicates, that the bound only applies under certain conditions. The new/additionally shown bounds here are from the $Z$-invisible width (for $m_{Z'}= 10\,\GeV$ only) and from parity violation using E158 under the assumption $g_\phi = g_L$ in the section  ``{ Precision Tests/LEP}''. Section 
     ``{Astro/Cosmo/Direct Detection}'' now shows the annihilation constraint from Voyager~1 data. The region of interest for $(g-2)_{e,\,\mu}$, where the bound for $(g-2)_{e}$ is based on \cite{Parker:2018vye,Morel:2020dww},  also requires further assumptions of couplings $g_\phi = g_L = \pm g_R$.  The INTEGRAL interpretation is excluded.}
    \label{Fig:summary_Z}
\end{figure*}

In light of the significant amount of activities in the past two decades that has gone into the exploration of the MeV-GeV mass range and the large amount of results, it seems  timely to revisit the originally proposed models of sub-GeV scalar DM~\cite{Boehm:2003hm} and confront them to this new wealth of data.  Concretely, we add the following new pieces that were not presented previously in this context:
\begin{itemize}\setlength\itemsep{0em} 
    \item sensitivities of current and future intensity-frontier experiments are derived for the first model, and re-visited for the second;
    \item in addition to an update of the $g-2$ constraint from electrons,  limits from lepton flavor violation, parity violation, and the invisible decay of the $Z$-boson are established;
    \item the astrophysical cooling constraint from SN1987A is derived in detail for both, the free-streaming and trapping regime;
    \item the bound from astrophysical DM self-scattering is derived, while we adopt the limits on DM annihilation at the CMB epoch and from Voyager~1 data at present time in the literature.
    \item the high-redshift constraint from the collisional damping of DM primordial fluctuations is considered and from extra radiation degrees of freedom is re-visited;
    \item latest constraints from the leading direct detection experiments are summarized to apply to the models;
    \item the sensitivity of the high-energy colliders, LEP and LHC,  is provided;
    \item an improved velocity expansion for the annihilation cross section for one of the models is presented.
\end{itemize}%
Taken together, this will provide a more comprehensive assessment as whether light dark matter particles could explain INTEGRAL or the anomalous magnetic moment of the muon in this setup.  It is important to stress, however, that although these anomalies serve as good and timely motivations, our study has the broader aspect that it presents a complete and self-contained survey on the viability of rather minimal models of scalar DM below the GeV-scale.  A summary of results for an exemplary DM mass of 10~MeV is shown in Figs.~\ref{Fig:summary_F} and~\ref{Fig:summary_Z}.

The paper is organized as follows: in Sec.~\ref{Sec:particle}  we  introduce the models together with the  parameter regions of central interest. In Sec.~\ref{Sec:lab} the bounds from the intensity-frontier experiments, from precision observables and from LEP are derived. While such bounds constrain parts of the parameter regions of interests, complementary limits arise from  cosmological and astrophysical observations, discussed in Sec.~\ref{Sec:cosmos}. Section~\ref{Sec:onshellZp} is devoted to a study of the low mediator mass region in the $Z'$ model. Conclusions are presented in Sec.~\ref{Sec:con}. Several appendices provide additional details on our calculations.

 \section{Status of the INTEGRAL-line}
 \label{status}
Before we go into the specifics of the scalar DM models, for self-completeness, in this section we provide a lighting review on the status of the INTEGRAL line in its connection to annihilating DM into $e^+e^-$ pairs. 
Decomposing the  annihilation cross section in terms of the relative velocity as $\sigma_{\rm ann} v = a + b v^2$ and assuming a NFW dark matter halo profile, the observations suggest that the best fit values for the  $a$ or $b$ parameters are~\cite{Boehm:2004gt}, 
\begin{subequations}
\label{Eq:INTEGRAL_values}
\begin{align}
	 a &\simeq 2.2\times 10^{-31}   \left( \dfrac{m_\phi}{\rm MeV}\right)^{2}\, {\rm cm}^3\, {\rm s}^{-1}\, \\
	 b & \simeq 3.4\times 10^{-25}   \left( \dfrac{m_\phi}{\rm MeV}\right)^{2}\, {\rm cm}^3\, {\rm s}^{-1}\,,
\end{align}
\end{subequations}
with a strong preference for a constant cross section ($a$-value) \cite{Ascasibar:2005rw} albeit large uncertainties and an additional dependence on the cuspiness of the inner DM halo profile~\cite{Vincent:2012an}. With these numbers in mind, the $p$-wave is roughly commensurate with the value required for a successful thermal relic $\sigma_{\rm ann} v \sim {\rm few}\times 10^{-26}\, {\rm cm^3}\,{\rm s^{-1}}$ (where $v\sim 0.3$ at freeze-out). 

The question of viable DM mass is an involved one. The injected positrons produced in DM annihilation need to decelerate to non-relativistic speeds before annihilating to explain the 511~keV INTEGRAL line. Because of substantial uncertainties in astrophysical propagation modelling~\cite{Higdon:2007fu,Lingenfelter:2009kx}, the maximal DM mass that can explain the line remains uncertain; see~\cite{Kierans:2019pkh} for a recent summary. 
There are, however, several quantitative results with regard to spectral features:

1) Extra photons created by Bremsstrahlung in the annihilation process suggest $m_\phi\lesssim 20\,\MeV$~\cite{Beacom:2004pe} although more detailed calculations relax this bound to $m_\phi\lesssim 30-100\,\MeV$~\cite{Boehm:2006df}. This is comparable to the Voyager~1 bound based on local $e^\pm$ measurements~\cite{Boudaud:2016mos}.

2) The most stringent constraint on the DM mass is obtained when considering the in-flight annihilation, implying $m_\phi\lesssim 3-7.5\,\MeV$~\cite{Beacom:2005qv,Sizun:2006uh}, mostly from the COMPTEL diffuse $\gamma$-ray background measurements. The constraint is derived from the X-ray background inside the gas-dense region of the Galactic center.

3) Recently, Ref.~\cite{Cirelli:2020bpc} has re-visited the extra photon emission (mainly via inverse Compton scattering) from DM annihilation at  higher latitudes, where in-flight annihilation is sub-leading. Such treatment leads to a much weaker bound from the INTEGRAL data, $m_\phi\lesssim 70\,\MeV$, when normalizing on the 511~keV line strength.
Furthermore, Ref.~\cite{Bartels:2017dpb} considers both  Bremsstrahlung and in-flight annihilation,  showing that a future e-ASTROGAM experiment is able to probe the DM mass down to 4~MeV.

\section{Representative models}
\label{Sec:particle}

In this paper we shall focus on a complex scalar DM candidate. The Galactic 511\,keV gamma ray line can then be explained by either $t$-channel or $s$-channel  annihilation processes~\cite{Boehm:2003hm}. The former process necessarily involves an electrically charged particle, taken as a fermion below. Without loss of generality, the $s$-channel case assumes the presence of an intermediate gauge boson, which we shall take as leptophilic.  

\subsection{Heavy fermion mediator \boldmath$F$}

In the first model that we consider, the scalar DM particle, denoted by $\phi$, couples to the Standard Model (SM) via heavy fermionic mediators. For the sake of generality we take $\phi$ to be complex, but mention applicable formul\ae\ for real $\phi$ along the way. 
Concretely, $\phi$ and its antiparticle~$\phi^*$ may couple to the SM charged and neutral leptons $l= (l^-_L, l^-_R)^T$ and $\nu_l$ through a Yukawa-like interaction with the introduction of new electrically charged and neutral fermions $F^\pm$ and $F^0$,  arranged as part of an SU(2)$_L$ doublet $(F_L^0,F^-_L)$, as well as  singlets  $F^0_R$ and $F^-_R$. Written in terms of Dirac fields $F=(F^-_L, F^-_R)^T$ and $F^0=(F_L^0,F_R^0)^T$, the Lagrangian reads,
\begin{align}
\label{Eq:Yukawa_int}
	\mathcal{L}_{F}	& =  - c^l_L \,\phi \bar F P_L l - c^l_R \phi \bar F P_R l - c^l_L \phi \bar F^0 P_L \nu_l + h.c.\, .
\end{align}
Here, $(\nu_l, l_L^-)$ and  $l^-_R$ are the SU(2)$_L$
doublets and singlets of lepton flavor $l = e, \mu, \tau$; $P_L = (1 - \gamma^5)/2$ and $P_R = (1 + \gamma^5)/2$ are the projection operators. We take all couplings to be real.
In the presence of right-handed neutrinos $\nu_R$, additional interactions become possible, 
\begin{equation}
\label{Eq:Yukawa_int_nu}
	\mathcal{L}'_{F} = - c^l_R \,\phi (\bar{F}_R^0 \nu_R ) + h.c.\,.
\end{equation}
For the purpose of this paper, we shall not consider the latter option in any detail, but mention applicable results in passing. 

There are a number of options  related to~\eqref{Eq:Yukawa_int}, see \textit{e.g.}~\cite{Agrawal:2011ze, Schmidt:2012yg, Zhang:2012da, Bai:2013iqa, Bai:2014osa, Agrawal:2014una, Kile:2014jea}. In what follows, we usually drop the superscript on $c^l_{L,R}$  for the coupling to electrons and electron-neutrinos as we consider them as always present, $c_{L,R}\equiv c^e_{L,R}$. 
Non-zero couplings to the second and third generations are {\it a priori} not the main focus of the paper, but they lead to further  interesting  consequences. Among them is  a contribution to the anomalous magnetic moment of the muon, discussed below. If there is a single generation of heavy fermions $F$---which is the way how the Lagrangian is written---one may additionally induce lepton-flavor violating processes between the electron sector and muon or tau sector for $c^\mu\neq 0$ or $c^\tau \neq 0$, respectively (see below). At the expense of considering three generations of heavy fermions, $F_l$, the flavor symmetry can be restored. Finally, we note that there is also a global dark U(1)- or $Z_2$-symmetry in~\eqref{Eq:Yukawa_int} between $\phi$ and $F$; the former (latter) applies for $F^0$ being Dirac (Majorana).

Because of collider bounds on  charged particles~\cite{Egana-Ugrinovic:2018roi}, the fermions $F$ have to be above the EW scale. Therefore, we take the advantage that they never appear on-shell in any process considered here, and 
derive constraints on the effective UV-scale  $\Lambda_F = (c_L c_R /m_F)^{-1}$. Before constraining the model, we infer the normalization points for the couplings from two particularly important predictions: the contribution to the anomalous magnetic lepton moment, as well as the DM annihilation cross section corresponding to the INTEGRAL signal and the thermal relic.

\subsubsection{Anomalous magnetic moment}

\begin{figure}
\begin{center}
\includegraphics[width=0.4\columnwidth]{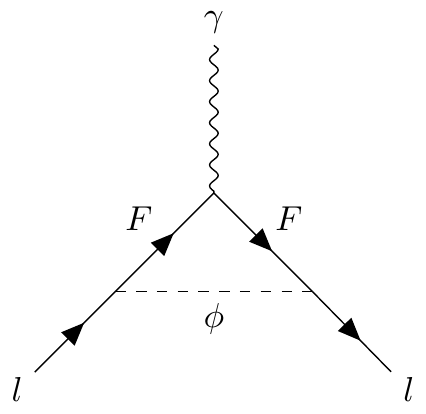}\hspace*{0.5cm}
\includegraphics[width=0.4\columnwidth]{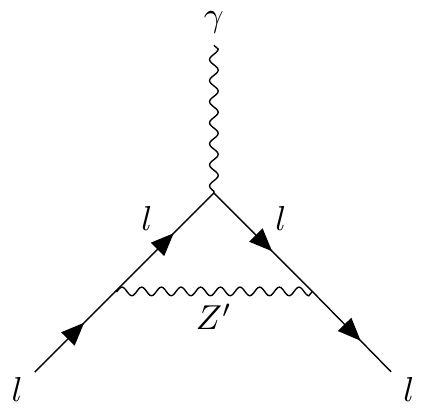}
\end{center}
\caption{Left panel: contribution to $(g-2)_l$ from $\phi$ and $F$ particles. Right panel: contribution to $(g-2)_l$  from new $Z'$ interaction.}
\label{Fig:gminus2}
\end{figure}

Under the assumption that $ m_F \gg m_\phi  \geq m_l$ and that all $c^l$-couplings are real, 
the one-loop contribution to the leptonic anomalous magnetic moment, shown in the left panel of Fig.~\ref{Fig:gminus2},  is given by 
\begin{equation}
\label{Eq:g-2_F}
	\Delta a_l^{(F)} = \dfrac{c^l_L  c^l_R}{16 \pi^2} \dfrac{m_l}{m_F},
\end{equation}
in agreement with previous calculations~\cite{Leveille:1977rc,Boehm:2003hm}; note that $a_l \equiv (g_l -2 )/2$.
Therefore, to address the long-standing muonic $g-2$ anomaly~\cite{Bennett:2006fi}
\begin{equation}
\label{Eq:gmu-2}
	  a_\mu^{\rm exp} - a_\mu^{\rm SM} = (290 \pm 90) \times 10^{-11},
\end{equation}
the corresponding favoured region is $c^\mu_F \equiv \sqrt{|c^\mu_L c^\mu_R|} \sim (5.5 - 7.6) \times 10^{-2}$ with $m_F = 1\,{\rm TeV}$.
The full expression without  assuming the mass hierarchy is given in App.~\ref{app:g-2}. In anticipation of the constraints to be derived below, we point out that the contribution~\eqref{Eq:g-2_F} to the electron anomalous magnetic moment will be of central importance when assessing the viability of explaining  various anomalies.

\subsubsection{DM annihilation}

In the model with heavy fermionic mediators $F^\pm$ and $F^0$, the non-relativistic DM annihilation cross section into $e^+e^-$ via $F^\pm$ exchange or into Dirac electron neutrinos via $F^0$ exchange with the participation of a (kinematically unsuppressed) light right-handed state given in Eq.~\eqref{Eq:Yukawa_int_nu}, $\bar \nu_e \nu_R$ or $\bar \nu_R \nu_e$,  reads%
\begin{equation}
\label{Eq:nonrel_annF_nsc}
  \sigma_{{\rm ann},F} v_M  =  \dfrac{ c_L^2 c_R^2  }{4 \pi m_F^2} \left(1 - \dfrac{m_l^2}{m_\phi^2} \right)^{\tfrac{3}{2}} + \, \dfrac{3c_L^2 c_R^2 m_l^2  v_{\rm rel}^2 }{32\pi m_F^2 m_\phi^2} \sqrt{ 1 - \dfrac{m_l^2}{m_\phi^2} } \,,
\end{equation}
where $v_{M} = 2(1-4m_\phi^2 /s)^{1/2}$ is the M{\o}ller velocity. The $s$-wave component agrees with the one in Eq.~(1) in~\cite{Boehm:2003hm}, while the $p$-wave component is different, due to the fact that we expand in the Lorentz invariant product $\sigma_{{\rm ann},F} v_{M} $ rather than  $\sigma_{{\rm ann},F} v_{\rm rel}$; see  App.~\ref{Sec:phi_pair_ann} for  the full expressions, as well as those for real scalar DM. Above we have omitted terms that are suppressed by  $(m_{l,\,\phi}/m_F)^4$ as well as higher-order terms. 
For the special case $c_L c_R =0$ and for $m_l  \to 0$ the above cross section vanishes, and the process becomes $d$-wave dominated~\cite{Toma:2013bka, Giacchino:2013bta, Giacchino:2014moa}, scaling as $v_{\rm rel}^4  m_\phi^6/m_F^8$. 
Given a TeV-scale $F$ and $m_{l,\,\phi}$ well below GeV-scale, the latter terms do not contribute to the annihilation cross section in any appreciable way.
 Finally, for real scalar $\phi$,  a factor of four should be multiplied to the expression in Eq.~\eqref{Eq:nonrel_annF_nsc}  as both $t$- and $u$-channel processes contribute.  

The annihilation to a pair of left-handed neutrinos $\bar\nu_l \nu_l$, mediated by  $F^0$, is either suppressed by neutrino mass or $1/m_F^4$ or $v^2/m_F^4$, for Dirac neutrinos; see Eqs.~\eqref{eq:annFlong} and~\eqref{eq:B2}.
However, if neutrinos are Majorana fermions, one may additionally annihilate to $\nu_l \nu_l$ or, equivalently, $\bar \nu_l \bar \nu_l$ with an $s$-wave cross section similar to~\eqref{Eq:nonrel_annF_nsc}~\cite{Boehm:2006mi}.  We comment on this possibility when considering cosmological constraints.

While we do not presume any production mechanism of the observed DM relic abundance, we will show the required parameters for  thermal freeze-out below. Here, the DM abundance is $\Omega_\phi h^2 = 0.1198$~\cite{Aghanim:2018eyx}, where $\Omega_\phi$ is the density parameter of $\phi$ and $h$ is the Hubble constant in units of 100~km/s/Mpc.
The observed relic density is achieved  with $c_F^2 \sim 0.01\text{--}0.1$ for  $m_\phi < \mathcal{O}({\rm GeV})$ and $m_F \sim \mathcal{O} (100\,{\rm GeV})\text{--}\mathcal{O} ({\rm TeV})$~\cite{Boehm:2003hm}.  The parameter regions that yield the required annihilation cross section within uncertainties for both the thermal freeze-out and the INTEGRAL $511\,{\rm keV}$ line~\cite{Wilkinson:2016gsy} are shown in Fig.~\ref{Fig:constraint_F} for $m_F \gg m_\phi$.

Although, as detailed in Sec.~\ref{status}, the 511\,keV line prefers a DM mass below several to tens of MeV, we scan over the entire MeV--GeV mass range as our results bear greater generality. In Figs.~\ref{Fig:constraint_F} and  \ref{Fig:constraint_Zp} below, we indicate by a lighter shading of the INTEGRAL favored bands the weakest constraint on $m_\phi$ that is derived from the INTEGRAL X-ray data itself~\cite{Cirelli:2020bpc}, $m_\phi \ge 70$\,MeV.

\subsection{Leptophilic vector mediator \boldmath$Z'$}

Turning now to the model  with a gauge boson $Z'$, both the DM particle $\phi$ and SM leptons are charged under the new U(1).
The interactions have the form 
\begin{align}
\label{eq:Lzprime}
     \mathcal{L}_{Z'} &= g_\phi^2 Z'_\mu Z'^\mu \phi^* \phi  -i g_\phi Z'_\mu \left[ \phi^*(\partial^\mu \phi ) - (\partial^\mu \phi^* ) \phi \right] \nonumber \\
    &\quad -  Z'_\mu  \bar{l} \gamma^\mu (g_L P_L + g_R P_R)l .
\end{align}
The couplings $g_{L,R}$ and $g_\phi$ are understood as a product of gauge coupling $g$ and charge assignments $q_{L,R}$ and $q_{\phi}$ so that $g_{L,R} = g q_{L,R} $ and $g_\phi = g q_\phi$, respectively. Again, there are many options available with~\eqref{eq:Lzprime}. They generally differ by the $Z'$ mass $m_{Z'}$, by their chiral couplings, by the absence or presence of family universality and/or kinetic mixing, by their (extended) Higgs sector, by potential additional fields that are required to cancel associated gauge anomalies in the UV and so forth; see e.g.~\cite{Langacker:2008yv} and references therein.

Here, we are primarily focused on the phenomenology associated with the $Z'$ coupling to electrons, and shall take $g_L$ and $g_R$ as flavor blind for when muons are involved. The special cases $g_l\equiv g_L=g_R$ and $g_L = -g_R$ correspond to a pure vector and axial-vector interactions, respectively.%
\footnote{For GeV fermionic DM with a leptophilic $Z'$, see \text{e.g.} \cite{Fox:2008kb, Bi:2009uj, Bell:2014tta}.}
For the purpose of illustration, we consider $m_{Z'} \ge 10\,$GeV in most of our discussions. As will be shown, only a $Z'$ below  the EW scale is of relevance for the INTEGRAL signal,  so appears on-shell at high-energy colliders. As a result, although $Z'$ is  generally off-shell for the low-energy phenomenology, and  bounds derived below can  be  represented using $\sqrt{g_\phi g_l }/m_{Z'}$, results from LEP need to be treated with caution. For the latter, we provide  bounds both on $\sqrt{g_\phi g_l }/m_{Z'}$ in the heavy mediator limit, and  on $\sqrt{g_\phi g_l}$ for $m_{Z'} \ll m_Z$.
The possibility of a $Z'$ below 10\,GeV will be  discussed separately  in Sec.~\ref{Sec:onshellZp}.

\subsubsection{Anomalous magnetic moment}

Similarly as above, for the case $m_{Z'} \gg m_l$, the one-loop contribution to $(g-2)_l$ is given by,
\begin{equation}
\label{Eq:g-2_Zp}
   \Delta a_l^{(Z')} =  \dfrac{6g_L g_R -2 (g_L^2+g_R^2) }{24\pi^2} \dfrac{m_l^2}{m_{Z'}^2}\,,
\end{equation}
in agreement with \cite{Boehm:2007na, Agrawal:2014ufa} if  a pure vector coupling $g_L = g_R \equiv g_l$ is assumed. 
The full expression of Eq.~\eqref{Eq:g-2_Zp} is found in App.~\ref{app:g-2}.
The associated diagram of interest is shown in the right panel of Fig.~\ref{Fig:gminus2},
and the $(g -2)_\mu$ favoured parameter space is $g_l/m_{Z'} \sim (4.6\text{--}6.4)\,$TeV$^{-1}$. The constraint from $(g-2)_e$ will be evaluated in Sec.~\ref{sec:loops}.

For a flavor-blind $g_l$ assumed here, the combination of several experiments excludes the possibility that this simple model explains the muonic $g-2$ anomaly. 
This conclusion holds irrespective of if $Z'$ decays dominantly into SM leptons or into DM particles, as the leading constraint comes from the measurements of electron-neutrino scattering~\cite{Bilmis:2015lja}.%
\footnote{Other $Z'$-options such as $U(1)_{L_\mu - L_\tau}$ remain allowed for resolving $(g-2)_{\mu}$, as most recently illustrated in~\cite{Cadeddu:2020nbr}. For bounds on other relevant DM models, see e.g. \cite{Agrawal:2014ufa}.}

\subsubsection{DM annihilation}

For the annihilation cross section via a $s$-channel $Z'$, the $s$-wave component vanishes as scalars have no spin, and the $p$-wave component reads
\begin{eqnarray}\label{Eq:nonrel_annZ}
    \sigma_{{\rm ann},Z'} v_M  &=&  v_{\rm rel}^2  \dfrac{4m_\phi^2 (g_L^2 +g_R^2 ) - m_l^2 (g_L^2 - 6 g_L g_R + g_R^2)}{48\pi   (m_{Z'}^2 - 4m_\phi^2)^2 }  \notag\\
    && \times \,  g_\phi^2  \sqrt{1 - m_l^2/m_\phi^2} \,. 
\end{eqnarray}
It agrees with Eq.~(3) in~\cite{Boehm:2003hm} when taking $v_{\rm rel} \simeq 2 v_\phi$, where $v_\phi$ is the DM velocity. Since the cross section only varies by about a factor of two when either $g_R =0$ or $g_L=0$,  we do not  distinguish the left- and right-chiral couplings any further for annihilation, and simple take $g_L = g_R \equiv g_l$ in the remainder. For real scalar DM, the annihilation would be extremely suppressed since the $Z'$ does not couple to a pair of real scalars at tree level. 

 Taking the DM annihilation $\phi\phi^* \to Z'^* \to  l^- l^+$ with cross section as above, for $m_{Z'} \gg m_\phi > m_l$, the parameter region of interest for  INTEGRAL is shown by the red and blue bands in  Fig.~\ref{Fig:constraint_Zp} for NFW and Einasto profiles, respectively.
Finally, we note that the observed DM relic abundance is  achieved when~\cite{Boehm:2003hm} 
 \begin{equation}
 \label{Eq:rd_Zp_Celine}
    g_\phi g_l \sim (3\text{--}12) \times \left( \dfrac{m_{Z'}}{\rm { 10\,GeV}}\right)^2 	\left( \dfrac{m_\phi}{\rm {MeV}}\right)^{-1}, 
 \end{equation}
in the limit of $ m_{Z'} \gg m_\phi $. Obviously, $m_{Z'}$ around or above the EW scale puts us into the non-perturbative regimes and is not of interest for us.  Depending on the $Z'$ decay width, resonant annihilation at the point $m_{Z'}\simeq 2 m_{\phi}$ introduces additional velocity dependence in the annihilation. A detailed investigation of the resonant point, such as performed in~\cite{Pospelov:2008jd, Ibe:2008ye, Cline:2017tka}, is beyond the scope of this paper. Due to these reasons, we focus on $m_{Z'} \ge 2.1 m_\phi $.

 \section{Laboratory constraints}
 \label{Sec:lab}
 
\begin{figure}
\begin{center}
\includegraphics[width=0.49\columnwidth]{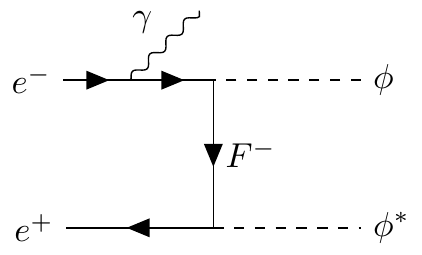}%
\includegraphics[width=0.49\columnwidth]{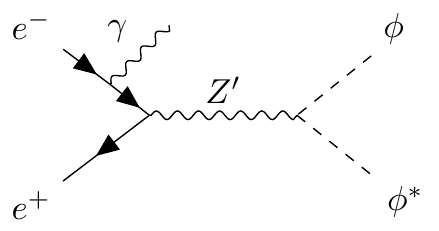}
\end{center}
\caption{
Pair production of $\phi$ in electron-positron annihilation in association with initial state radiation. Photon emission from the intermediate charged $F$ is suppressed and hence neglected.}
\label{Fig:ecollider}
\end{figure}

\begin{figure}
\begin{center}
\includegraphics[width=0.49\columnwidth]{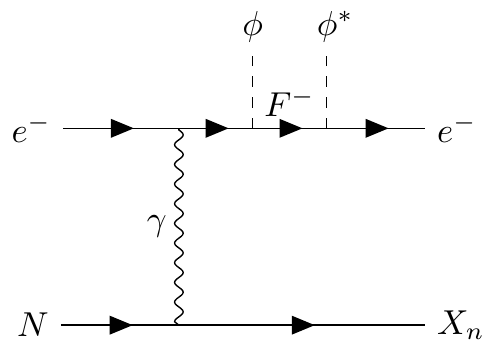}%
\includegraphics[width=0.49\columnwidth]{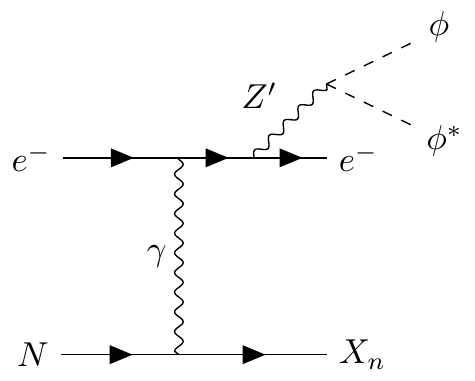}
\end{center}
\caption{
Pair production of $\phi$ in electron-beam fixed-target experiments. We consider $\phi$ emission from both initial and final state electrons (but not from the heavy $F$ particle). 
Note that a global dark symmetry in Eq.~\eqref{Eq:Yukawa_int} forbids the diagram with $\phi$ and $\phi^*$ interchanged for the left process; see main text.
}
\label{Fig:fixed-target}
\end{figure}

 The dark sector particles  may be produced in the  laboratory, especially at electron-beam facilities, through electron-positron annihilation in colliders  (Fig.~\ref{Fig:ecollider}) or electron-nuclei bremsstrahlung in  fixed-target experiments (Fig.~\ref{Fig:fixed-target}). Moreover, they can also appear virtually through loops, affecting EW precision measurements. Such considerations thus put upper bounds on the coupling of SM particles to the dark sector.  
 
 We briefly introduce the experimental data of interest and our methods to derive the related constraints below, and refer to \cite{Chu:2018qrm} and our appendices for further details of relevant cross sections.

\subsection{Electron-beam facilities}

We first consider intensity frontier experiments, including  low-energy  electron-positron colliders and electron-beam fixed-target experiments. 
For the  values of $m_F$ and $m_{Z'}$ concerned above, we can only produce $\phi$ via off-shell mediators in these experiments.

Following our previous work~\cite{Chu:2018qrm},  we derive the expected number of signal events and constraints from current experiments such as BaBar~\cite{Aubert:2001tu}, NA64~\cite{Banerjee:2017hhz,Gninenko:2719646} and mQ~\cite{Prinz:1998ua,Prinz:2001qz}, as well as projected sensitivities for future ones, including Belle II~\cite{Abashian:2000cg, Abe:2010gxa}, LDMX~\cite{Mans:2017vej} and BDX~\cite{Battaglieri:2017qen}. Depending on the observable signatures, these experiments can be put in three categories described below. 

The first category is to look for large missing transverse momentum/energy, accompanied by a mono-photon signal, in low-energy electron-positron colliders, such as  BaBar and Belle II.  
The expected number of signal events in each energy bin reads
\begin{equation}
    N_{\rm sig}^{(i)} = \epsilon_{\rm eff} \mathcal{L} \int_{{\rm bin},i} \dfrac {ds_{\phi \phi}}{s} \int_{\cos\theta_\gamma^{\rm min}}^{\cos\theta_\gamma^{\rm max}} d\cos\theta_\gamma \dfrac{d\sigma_{e^- e^+ \rightarrow \phi\phi\gamma}}{dx_\gamma \, d\cos\theta_\gamma}\,,
\end{equation}
where $\epsilon_{\rm eff}$ is the efficiency, $\mathcal{L}$ is the integrated luminosity, $\theta_\gamma^{\rm max,min}$ are the cuts on the photon angle in centre-of-mass (CM) frame w.r.t. the beam axis, $x_\gamma = E_\gamma/ \sqrt{s}$ is the energy fraction carried away by the photon with $s$ and $s_{\phi\phi} = (1-x_\gamma) s$ being the CM energy square of $e^- e^+$ and $\phi$-pair, respectively.
The differential cross section is found in App.~\ref{Sec:epem_collider_xsec}.
For BaBar, we take the data of the analysis of mono-photon events in a search for invisible decays of a light scalar at the $\Upsilon (3{\rm S})$ resonance~\cite{Aubert:2001tu}. The CM energy is $10.35 \,{\rm GeV}$, with two search regions of $3.2\,{\rm GeV}\le E_\gamma \le 5.5\,{\rm GeV}$ and $2.2\,{\rm GeV}\le E_\gamma \le 3.7\,{\rm GeV}$.\footnote{We do not consider possible resonant conversion of $\Upsilon (3{\rm S}) \to Z'$ plus a low-energy photon, as $m_{Z'} = 10$\,GeV is chosen arbitrarily.}  
For Belle II, we follow~\cite{Essig:2013vha} and derive the projection by scaling up the BaBar background to an integrated luminosity of $50 \,{\rm ab}^{-1}$ with a similar CM energy.%

The second category describes the missing energy search in electron-beam fixed-target experiments, such as NA64 and future LDMX. 
The expected number of signal single-electron events is given by 
\begin{align}
        N_{\rm sig} &=N_{\rm EOT} \dfrac{\rho_{\rm target}}{m_N} X_0 \int^{E_{\rm max}}_{E_{\rm min}} dE_3 \,  \epsilon_{\rm eff} (E_3)\nonumber \\& \times \int^{\cos\theta_3^{\rm max}}_{\cos\theta_3^{\rm min}} d\cos\theta_3 \dfrac{d\sigma_{2\rightarrow 4}}{dE_3\,d\cos\theta_3}\,
\end{align}
in the thin target limit, where $N_{\rm EOT}$ is the number of electrons on target (EOT), $\rho_{\rm target}$ is the mass density of the target, $m_N$ is the target nuclei mass, $X_0$ is the radiation length of the target. $E_3$ is the energy of the final state electron and $\theta_3$ is the scattering angle w.r.t. the beam axis of the final state electron in the lab frame, with its detection efficiency given by $\epsilon_{\rm eff} (E_3)$. 
The differential cross section is derived in App.~\ref{Sec:eNbrem}.
The background in such experiments is usually negligible after imposing stringent selection criteria. %
The  NA64 experiment uses an electron beam with $E_{\rm beam} = 100\,{\rm GeV}$ and  has collected data of $N_{\rm EOT} = 4.3 \times 10^{10}$.
We select events with only a final state electron, with its energy between $[0.3,  50]\,{\rm GeV}$ and $\theta_3 \leq 0.23\,$rad.
For the proposed LDMX experiment, we use the benchmark values of phase I with $N_{\rm EOT} = 4\times 10^{14}$ at $E_{\rm beam} = 4\,{\rm GeV}$ and phase II with $N_{\rm EOT} = 3.2\times 10^{15}$ at $E_{\rm beam} = 8\,{\rm GeV}$. The energy and geometry cuts on final state electrons are $50\,{\rm MeV} < E_3 < 0.5 E_{\rm beam}$ and $\theta_3 < \pi /4$.  A constant $\epsilon_{\rm eff} =0.5$ is taken for both experiments.

The last category includes mQ and BDX, which are electron-beam fixed-target experiments designed to directly observe $\phi$-$e$ (or $\phi$-nucleon) recoil events  in a downstream detector. 
The expected number of electron recoil events is given by
\begin{equation}
    N_{\rm sig} = n_e L_{\rm det} \int_{m_\phi}^{E_\phi^{\rm max}} \int_{E_R^{\rm th}}^{E_R^{\rm max}} dE_R \, \epsilon_{\rm eff} (E_R) \dfrac{dN_\phi}{dE_\phi} \dfrac{d\sigma_{\phi \text{-}e}}{dE_R}\,,
\end{equation}
where $n_e$ is the electron number density in the detector, $L_{\rm det}$ is the detector depth.  The threshold recoil energy $E_R^{\rm th}$ depends on the experiment and $E_R^{\rm max}$ reads
\begin{equation}
  E_R^{\rm max} = \dfrac{2m_e (E_\phi^2 - m_\phi^2)}{m_e (2E_\phi + m_e) + m_\phi^2}\,,
\end{equation}
with the exact differential recoil cross section given in App.~\ref{Sec:scattering}.
The production spectrum of $\phi$ is computed by
\begin{align}
        \dfrac{dN_\phi}{dE_\phi} &= 2 N_{\rm EOT} \dfrac{\rho_{\rm target}}{m_N} X_0 \int_{E_\phi}^{E_{\rm beam}} dE \nonumber \\ &\times \int_{\cos\theta_\phi^{\rm min}}^{\cos\theta_\phi^{\rm max}} d\cos\theta_\phi\,I (E) \dfrac{d\sigma_{2 \rightarrow 4}}{dE_\phi\,d\cos\theta_\phi}\,,
\end{align}
in which the factor $2$ accounts for the production of the $\phi$-pair, $\theta_\phi$ is the scattering angle w.r.t.~the beam axis of the produced $\phi$ in the lab frame with boundaries given by the geometry of the downstream detector and $I(E)$ is the integrated energy distribution of electrons during their propagation in the target~\cite{Chu:2018qrm}.
The differential cross section for the $\phi$ energy distribution is listed in App.~\ref{Sec:eNbrem}.
For the mQ experiment, the incoming electron with energy $E_{\rm beam} = 29.5\,{\rm GeV}$ impinges upon a tungsten target with $N_{\rm EOT} = 8.4\times 10^{18}$. The collaboration has  reported 207 recoil events above the background, which is below the  uncertainty of the latter $\sigma_{\rm bkg} = 382$ within the signal time window.  We derive the upper bounds on the dark sector couplings from events with electron recoil energy $E_R \geq 0.1\,{\rm MeV}$. 
For  BDX, electrons with $E_{\rm beam} = 11\,{\rm GeV}$ are incident on an aluminium target which comprises 80 layers with thickness of $1$--$2\,{\rm cm}$ each. The BDX collaboration estimated that for $N_{\rm EOT} = 10^{22}$ the number of background events with $E_R\ge 0.35 \,{\rm GeV}$ is about 4.7~\cite{Battaglieri:2019nok}.  Again, we only consider electron recoil events, with a detection efficiency of $100\%$ for mQ~\cite{Prinz:2001qz}, and $20\%$ for BDX~\cite{Battaglieri:2019nok}.

 \begin{figure*}[t]
    \centering
    \includegraphics[width=0.45\linewidth]{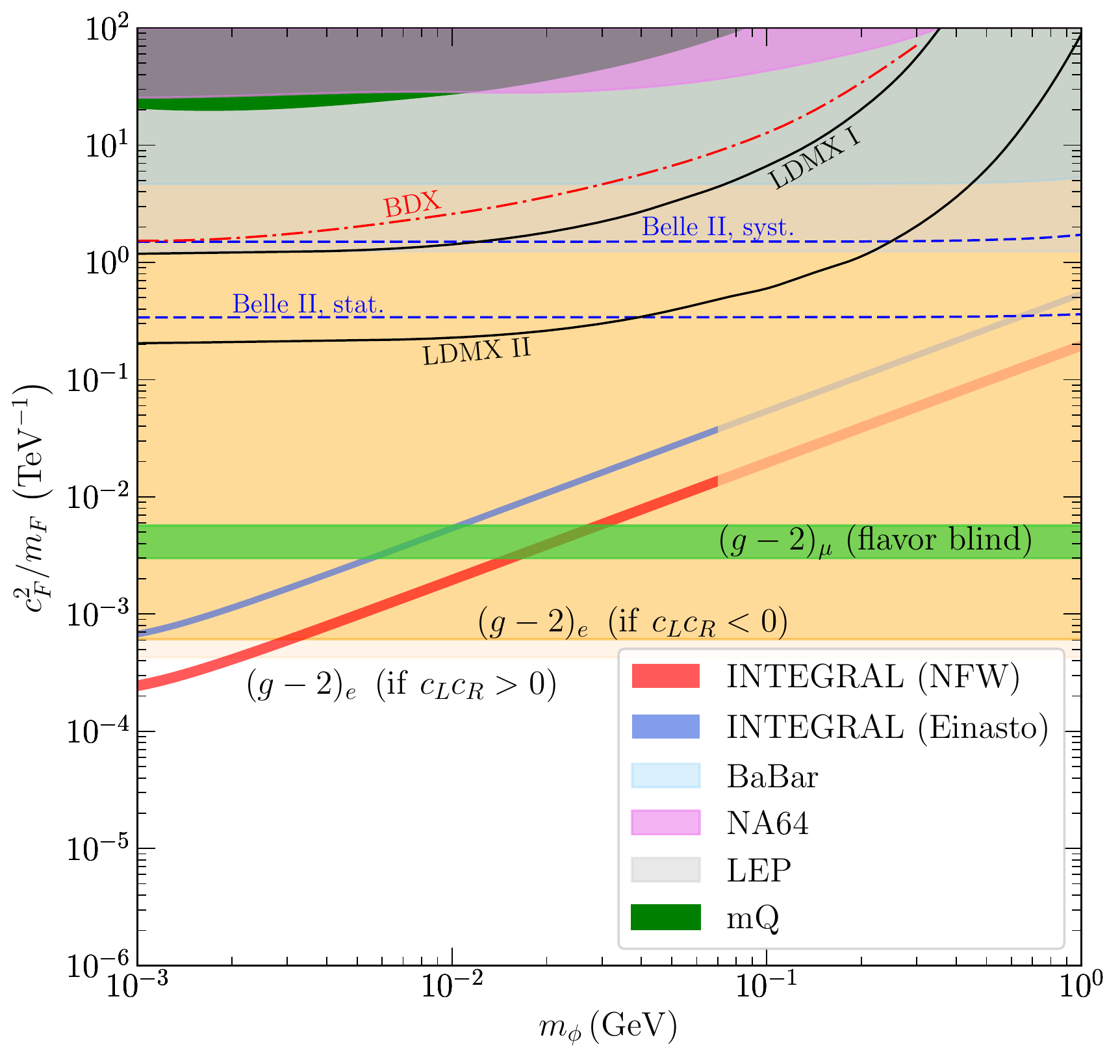}
     \includegraphics[width=0.45\linewidth]{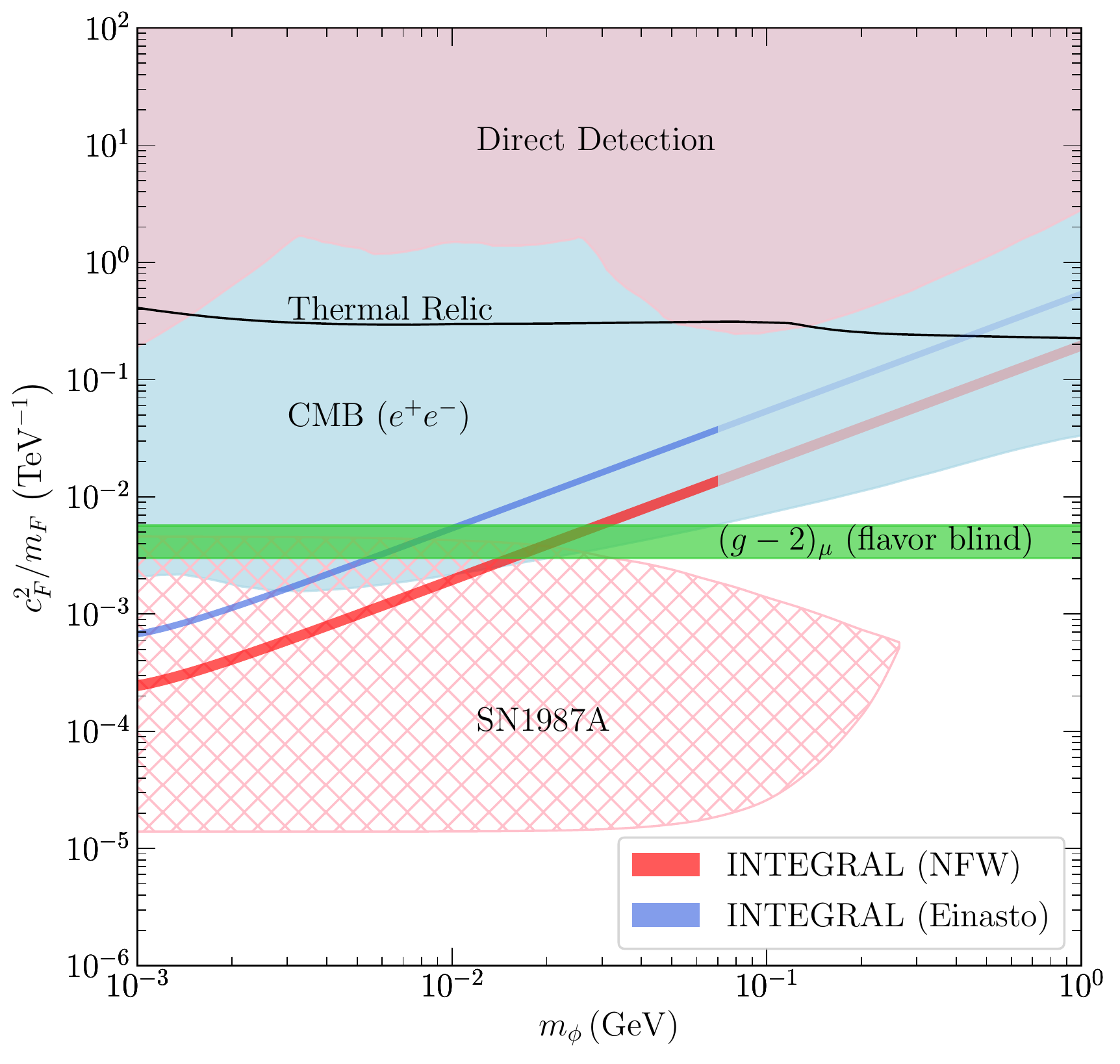}
    \caption{Bounds on the inverse of effective UV-scale $\Lambda_F^{-1} = c_F^2/m_F$ in the $F$-mediated model from laboratory experiments (left panel) and from astrophysical observations including direct detection (right panel). The parameter regions of interest for the INTEGRAL excess are shown as thin blue and red bands; for $m_\phi \ge 70\,$MeV the DM interpretation is disfavored  from INTEGRAL itself~\cite{Cirelli:2020bpc} as indicated by a lighter shading. The green horizontal band where  $(g -2)_\mu$ is explained carries the assumption  $c^\mu_F = c_F^e$.  The $(g-2)_e$ constraint is the conservative one based on~\cite{Parker:2018vye,Morel:2020dww}, for both $c_L^e c_R^e <0 $ and $c_L^e c_R^e >0$. }
    \label{Fig:constraint_F}
\end{figure*}

 \subsection{High-energy colliders}
 
High-energy colliders may produce any of the dark sector particles studied here, leading to missing energy signatures. 
In the $F$ model, a TeV-mass charged fermion $F$ remains largely unconstrained by current bounds from LEP or LHC data, while the missing energy search in LEP~\cite{Achard:2003tx} is able to constrain the overall coupling as
\begin{equation}
 c_F^2/m_F \lesssim 1.23\,{\rm TeV}^{-1}\,,
\end{equation}
which can be improved by investigating DM production via Drell-Yan processes with high-luminosity LHC~\cite{Bai:2014osa}, as well with ILC~\cite{Rossi-Torres:2015eua}.  

For the $Z'$ model, the LEP bound varies depending on the  $Z'$ mass. For a heavy $Z'$ above the LEP energy scale, we obtain a bound from missing energy events induced by DM pair production as
\begin{equation}\label{eq:ZpLEPoffshell}
 \sqrt{g_\phi g_l}/m_{Z'}   \lesssim 2.89\,{\rm TeV}^{-1} \,,
\end{equation}
in agreement with previous results~\cite{Fox:2011fx, Cheung:2012gi}, which is stronger than the reach of low-energy beam experiments. Although this is shown in Fig.~\ref{Fig:constraint_Zp}, it does not apply to  $m_{Z'}$ below the LEP energy scale, where a more proper LEP bound may come from missing energy induced by on-shell $Z'$ production, requiring  $g_l \lesssim 0.01$~\cite{Ilten:2018crw}.  Its combination with the perturbative condition $ g_\phi^2/(4\pi) \lesssim 10$ results in
\begin{equation}\label{eq:ZpLEP}
 \sqrt{g_\phi g_l}   \lesssim  0.335\,,
\end{equation}  
being comparable to the BaBar bound for $m_{Z'} = 10\,$GeV. Naively speaking, these two LEP bounds, valid for different parameter regions of $m_{Z'}$, converge at $m_{Z'} \sim m_Z$. Projected sensitivities on a leptophilic $Z'$ portal have also been derived for future colliders, see \cite{Freitas:2014jla, Freitas:2014pua, delAguila:2014soa, Bell:2014tta, Chen:2015tia}.

\subsection{Precision observables induced by loops}
\label{sec:loops}

The measurement of the fine structure constant, $\alpha$, has been improved significantly with Cs atom interferometers~\cite{Parker:2018vye}. Taking as input  $\alpha \equiv \alpha^{({\rm Cs})}$, the SM prediction of the electron anomalous magnetic moment $  a_e^{\rm (Cs)} = a_e^{\rm SM}(\alpha^{({\rm Cs})})$ is now in  2.5$\sigma$ tension with the direct measurement of $a_e$~\cite{Hanneke:2008tm},  $a_e^{(\rm meas.)}- a_e^{\rm (Cs)} = - 0.88 (0.36)\times 10^{-12} $. At face value, this puts a  stringent requirement on a
new physics contribution, 
\begin{align}
\label{eq:ge}
   \left. \Delta a^{\rm BSM}_e\right|_{\rm Cs} & \in (- 0.88 \,\pm\, 3 \times 0.36 )\times 10^{-12}\nonumber \\ & = [ -1.96 ,0.20]\times 10^{-12}\, 
\end{align}
with a nominal $3\sigma$ requirement.  Most recently, another experiment, using the recoil velocity on a Rb atom, has  measured the value of the fine-structure constant with similar uncertainty~\cite{Morel:2020dww}.  Its value of~$\alpha$ suggests a smaller~$a_e$, in better agreement with the direct measurement, $a_e^{(\rm meas.)}- a_e^{\rm (Rb)} =  0.48 (0.30)\times 10^{-12} $. From this we can  obtain a similar constraint on the new physics contribution  
\begin{equation}
\label{eq:ge_new}
   \left. \Delta a^{\rm BSM}_e\right|_{\rm Rb} \in   [ -0.42 ,1.38]\times 10^{-12}\, .
\end{equation}
These differences above  could also be rephrased in tensions between $\alpha$ extracted from Cs/Rb experiments and from direct $a_e$ measurements using the standard model prediction, $\alpha(a_e^{\rm SM})$, i.e. in $\alpha^{\rm (Cs/Rb)}- \alpha(a_e^{\rm SM}) $. Both models---through their contribution to $a_e$---then imply an inferred shift in the value of $\alpha$. One should obtain the same constraints from both.

In the $F$ model, positive (negative) $c_L c_R$  yields a positive (negative) contribution; cf.~\eqref{Eq:g-2_F} or the full expression in App.~\ref{app:g-2}. 
As shown in Fig.~\ref{Fig:constraint_F}, either sign  then puts a  strong constraint on the  model with a $F$ mediator.
In the $Z'$ model, $g_L = g_R$ and $g_L = -g_R$ can also give a distinct contribution to $a_e$; see Eq.~\eqref{Eq:g-2_Zp}.
A conservative limit can be given by combining the weaker of each limits of~\eqref{eq:ge} and~\eqref{eq:ge_new}, {\it i.e.}~the lower bound from $\left. \Delta a^{\rm BSM}_e\right|_{\rm Cs}$ and the upper bound from  $\left. \Delta a^{\rm BSM}_e\right|_{\rm Rb} $. This yields
\begin{align*}
  4.3 \times 10^{-4}\,{\rm TeV}^{-1}&\ge c_Lc_R/m_F \ge -6.1\times 10^{-4}\,{\rm TeV}^{-1}\,, \\
  625\,{\rm TeV}^{-2} &\ge g_L g_R/m_{Z'}^2 \ge -178\,{\rm TeV}^{-2}\,, 
\end{align*}
for the $F$ and $Z'$ model. In contrast, the combination of the stronger limits results  in  
\begin{align*}
  6.2 \times 10^{-5}\,{\rm TeV}^{-1} &\ge c_Lc_R/m_F \ge -1.3 \times 10^{-4}\,{\rm TeV}^{-1} \,, \\
  91\,{\rm TeV}^{-2} &\ge g_L g_R/m_{Z'}^2 \ge -38 \,{\rm TeV}^{-2}\,.
\end{align*}
For the $Z'$ model, the $(g-2)_e$ constraint is always surpassed by the LEP bound above, for both $m_{Z'}\gg m_{Z}$ and $m_{Z'}\le m_{Z}$, and is hence not included in Fig.~\ref{Fig:constraint_Zp}; see Fig.~\ref{Fig:summary_Z}.\footnote{Another observable is the running of the fine structure constant, given by the photon vacuum polarization induced by the charged $F$-loop, $\Pi(-M_Z^2)-\Pi(0)$.  This number needs to be below $0.00018$~\cite{Tanabashi:2018oca}, requiring $m_F \gtrsim 80\,$GeV.
The formula for $\Pi(p^2)$ is given in Eq.~\eqref{Eq:selfenergy} with $g_l$ replaced by $e$. A dark U(1) gauge boson $Z'$ does not contribute to the running at one-loop.
}

One may exercise some caution  if exclusively applying~\eqref{eq:ge}, as it takes a positive half-$\sigma$ shift to rule out any model by increasing the $2.5\sigma$ tension to $3\sigma$. Here we stress that both the $F$- and $Z'$-mediated models allow for both signs in their contributions. Therefore, going in the other direction, one may first bring both measurements into reconciliation and in a further consequence, allow for a particularly large shift before the lower boundary in~\eqref{eq:ge} is reached. In this sense,~\eqref{eq:ge} entails both, an aggressive and conservative limit. In Fig.~\ref{Fig:constraint_F} we show  the conservative limits that arise from the combination of~\eqref{eq:ge} and~\eqref{eq:ge_new}.

The invisible decay $Z \rightarrow \phi\phi^*$ induced by the 1-loop diagram containing $F$ or $Z'$ will alter the decay width of $Z$, see Fig.~\ref{Fig:charged_Floop}. Such additional contribution is bounded by experiments~\cite{Zyla:2020zbs} to satisfy 
\begin{equation}
\label{Eq:Z_inv_bound}
   \Gamma ( Z \rightarrow {\rm inv} )_{\rm new} \lesssim 0.56 \,{\rm MeV}\quad {\rm at} \,\, 95\% \,{\rm C.L.}\, .
\end{equation}
Explicit calculation of the relevant loop diagrams, detailed in App.~\ref{app:zwidth}, reveals that the ensuing constraints ($c_F/m_F < 26.6$\,TeV$^{-1}$ and $\sqrt{g_\phi g_l} < 0.35$ for $m_{Z'} = 10\,$GeV) are  
 weaker than those above from general missing energy searches. We hence do not show this constraint in the figures.

\subsection{Further precision tests}

Before closing this section, we also mention that further limits arise when the models introduce new sources of parity or flavor violation. 

The effect of parity violation can be parametrized as a deviation from the SM-predicted value of the weak angle~$\theta_W$.%
\footnote{Here we note that there is no sensitivity from the feats that detected atomic parity violation~\cite{Bouchiat:1997mj}, because of the leptophilic nature of couplings involved; see~\cite{Bouchiat:2004sp,Fayet:2016nyc} for when a $Z'$ coupling to quarks is present.} The E158 experiment at SLAC used polarized electron beams of $E_e \simeq 46\text{--}48\,{\rm GeV}$, and measured the M\o{}ller scattering asymmetry with one polarized electron
\begin{equation}
A_{\rm LR}={ d\sigma_{e_{\rm R} e } - d\sigma_{e_{\rm L} e } \over d\sigma_{e_{\rm R} e } +  d\sigma_{e_{\rm L} e }} 
\end{equation}
at low momentum-transfer $Q^2 \simeq 0.026$\,GeV$^2$~\cite{Czarnecki:2000ic, Anthony:2005pm}. The subtracted value of $4\sin^2\theta_W -1$ from the data is $-0.0369(52)$,  slightly higher than the SM prediction, $ - 0.0435(9)$ at low energy.  For the $Z'$ model with $m_{Z'}\gg \sqrt{Q^2}$, the resulting bound scales as $g_{R(L)}/m_{Z'} \lesssim 0.38\,(0.58) $\,TeV$^{-1}$ at $g_{L(R)}=0$ and almost vanishes at $g_L = \pm g_R$. Moreover, at $g_L = g_R\cos2\theta_W /(\cos2\theta_W-1)$, the contribution of $Z'$ can be  absorbed by re-scaling the Fermi constant. Although for $m_{Z'} \ge 10$\,GeV concerned in this section, it is at most comparable with the LEP bound  above, future experiments, such as MOLLER~\cite{Benesch:2014bas} and P2 at MESA~\cite{Becker:2018ggl}, have the potential to improve the limit by more than one order of magnitude~\cite{Kumar:2013yoa,Arcadi:2019uif}. In contrast, the $F$ model does not induce additional electron scattering processes at tree level, and thus can hardly be constrained by such experiments. 

Moreover, if only one generation of $F$ is present, non-zero couplings to the muon and tau sector induce lepton flavor violation, similar to flavored DM~\cite{Agrawal:2011ze}. For example, one may have the decay $\mu \to e \gamma$ by closing the $\phi$-loop, effectively via the magnetic dipole interaction~\cite{Kuno:1999jp}. The current strongest limit, from  the MEG experiment~\cite{TheMEG:2016wtm}, requires $\text{Br}_{\mu \to e\gamma} \lesssim 4.2\times 10^{-13}$. This in turn gives $c_F^\mu c_F^e /m_F \lesssim 2.7 \times 10^{-5}\,$TeV$^{-1}$ if $c^l_L =c^l_R$ with $ c^l_F \equiv \sqrt{|c^l_L  c^l_R|}$. For purely chiral interaction with $c^l_{L}=0$, the bound is relaxed to $\sqrt{|c_{R}^\mu c_{R}^e|} /m_F \lesssim 0.05\,$TeV$^{-1}$ as an additional spin-flip is needed~\cite{Klasen:2016vgl}; the same bound applies if one switches the chiralty subscripts. 

Another example is the decay $\mu\to e\phi \phi $ at a rate compared to the SM mode, 
\begin{align}
    \frac{\Gamma_{e \phi\phi}}{\Gamma_{e\bar\nu_e\nu_\mu}} \sim \frac{m_W^4}{m_\mu^2 m_F^2} \quad {\rm or}\quad  \frac{m_W^4}{m_F^4}\,.
\end{align}
The first scaling applies 
for $(c^\mu_L c^e_R)^2 + (c^\mu_R c^e_L)^2 \neq 0$; the second scaling applies for when the same combination of couplings vanishes. Hence, because of the coupling structure and because of kinematic effects, one generally expects distortions of the electron spectrum. The latter is an important test for the $V-A$ nature of weak interactions and has been mapped out well in the coefficients describing it~\cite{Zyla:2020zbs, Renga:2019mpg}. The sensitivity is, however, likely superseded by the radiative decay above, and only applies to $m_\phi \le m_{\mu}/2$; a detailed study of it is beyond the scope of this paper.

Finally, we mention that for a $Z'$ boson that couples to quarks/leptons with appreciable strength,  precision observables were also investigated in~\cite{Davoudiasl:2012ag, DEramo:2017zqw}; note however that stringent constraints from dilepton resonance searches derived in the latter work are avoided, as in our setup $Z'$ dominantly decays into a $\phi$-pair. 

\section{Cosmological and astrophysical constraints }
\label{Sec:cosmos}

 As the scalar $\phi$ is assumed to be the dominant DM component, the models are also constrained by astrophysical and cosmological observables, as well as from DM direct detection experiments. These constraints are discussed in the following.   

 \subsection{BBN/CMB \boldmath$\Delta N_{\rm  eff}$ bounds}

Here we take into account the BBN/CMB bounds on $N_{\rm eff}$ from early Universe observations, while at the same time remaining agnostic about the state of the Universe for $T\gtrsim \MeV$. Since $m_\phi \sim \mathcal{O}({\rm MeV})$, during Big Bang Nucleosythesis (BBN) $\phi$ can still be relativistic and contribute to the radiation density, summarized in the parameter $N_{\rm eff}$. Recall that we always set $m_{Z'} \ge 2.1m_\phi$, so $Z'$ only plays a sub-leading role in the radiation density budget, even though it has three degrees of freedom.  Currently,  two relativistic degrees of freedom, like from a thermal complex scalar, are still considered to be  marginally allowed by BBN on $\Delta N_{\rm eff}$~\cite{Steigman:2010pa,2011PhLB..701..296M,Hamann:2011ge}.  

In contrast, the Planck measurement of the cosmic microwave background (CMB) spectrum requires that $N_{\rm eff}= 2.99\pm 0.33$ at the last scattering surface~\cite{Aghanim:2018eyx}. This limits the residual DM annihilation after neutrino-decoupling that injects energy either into the visible  or into the neutrino sector~\cite{Boehm:2013jpa, Heo:2015kra}. In the $F$ model,  $\phi$ pairs annihilate into electrons.  Under the assumption of a sudden neutrino decoupling at $1.41$\,MeV~\cite{Fornengo:1997wa}, we obtain a lower bound from $N_{\rm eff}$ as $m_\phi \ge 5.1$\,MeV  for a complex  scalar, consistent with previous results~\cite{Wilkinson:2016gsy}.%
\footnote{In the case of Dirac SM neutrinos with a kinematically accessible right-handed neutrino (as alluded to when introducing the models), one also would need to verify that $\nu_R$ decouples early enough from the thermal bath, so that, overall, the upper bound on $N_{\rm eff}$ is satisfied.} However, the CMB bounds from  $N_{\rm eff}$ become much weaker if  the scalar DM annihilates into both electrons and neutrinos, which happens in the $F$ model with Majorana neutrinos, as well as in the $Z'$ model.
The underlying reason is that both, the photon- and neutrino-fluid is being heated so that the ensuing offset in the ratio of their respective temperatures is milder;  see e.g. \cite{Escudero:2018mvt, Depta:2019lbe, Sabti:2019mhn} for recent discussions.
In a flavor-blind set-up assuming  DM annihilates to electrons and each species of SM neutrinos equally, we then estimate that the Planck bound on $N_{\rm eff}$ only requires $m_{\phi} \gtrsim 2.0$\,MeV. The latter possibility was not considered in~\cite{Wilkinson:2016gsy}.

 \begin{figure*}[t]
    \centering
    \includegraphics[width=0.45\linewidth]{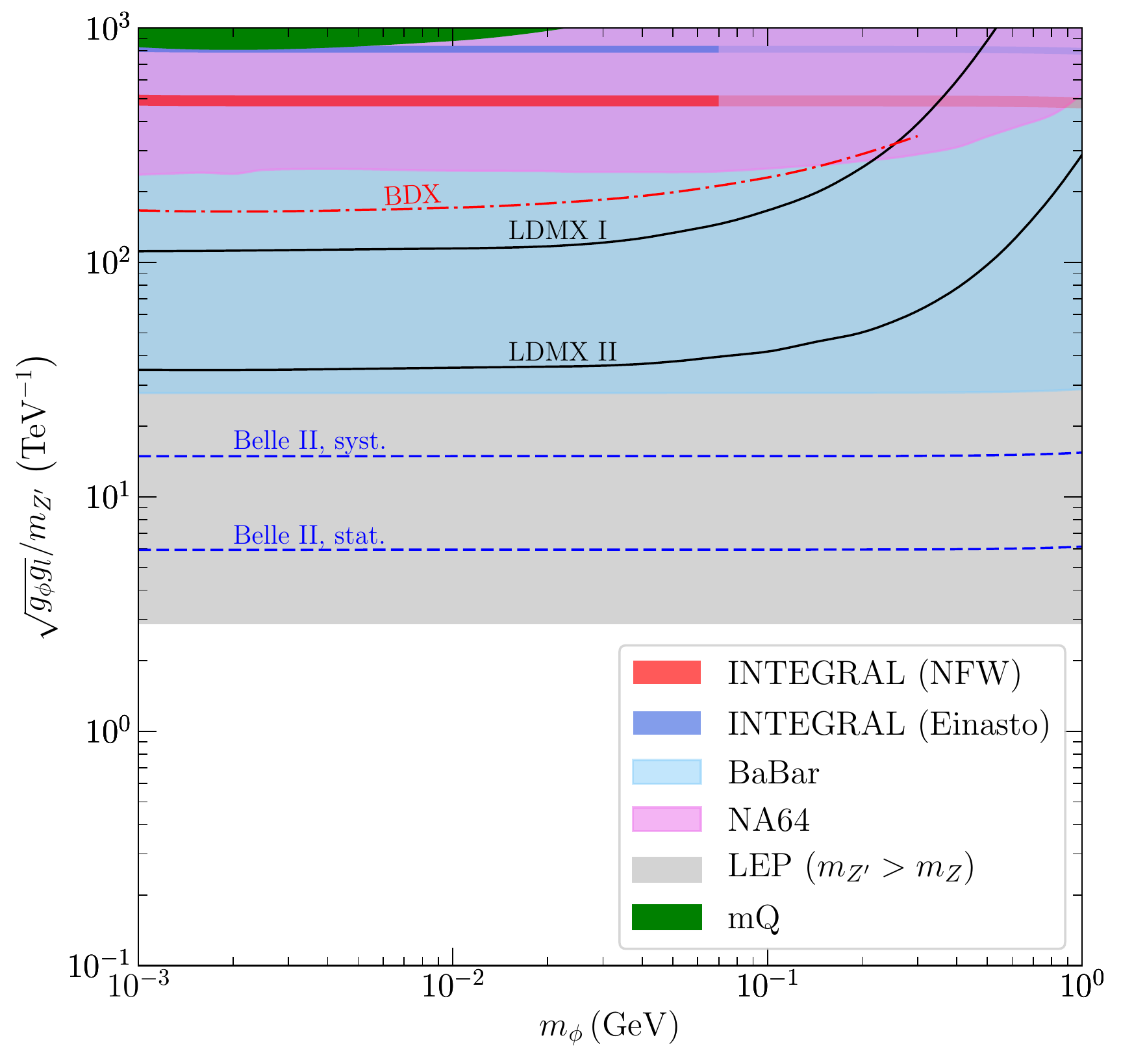}
    \includegraphics[width=0.45\linewidth]{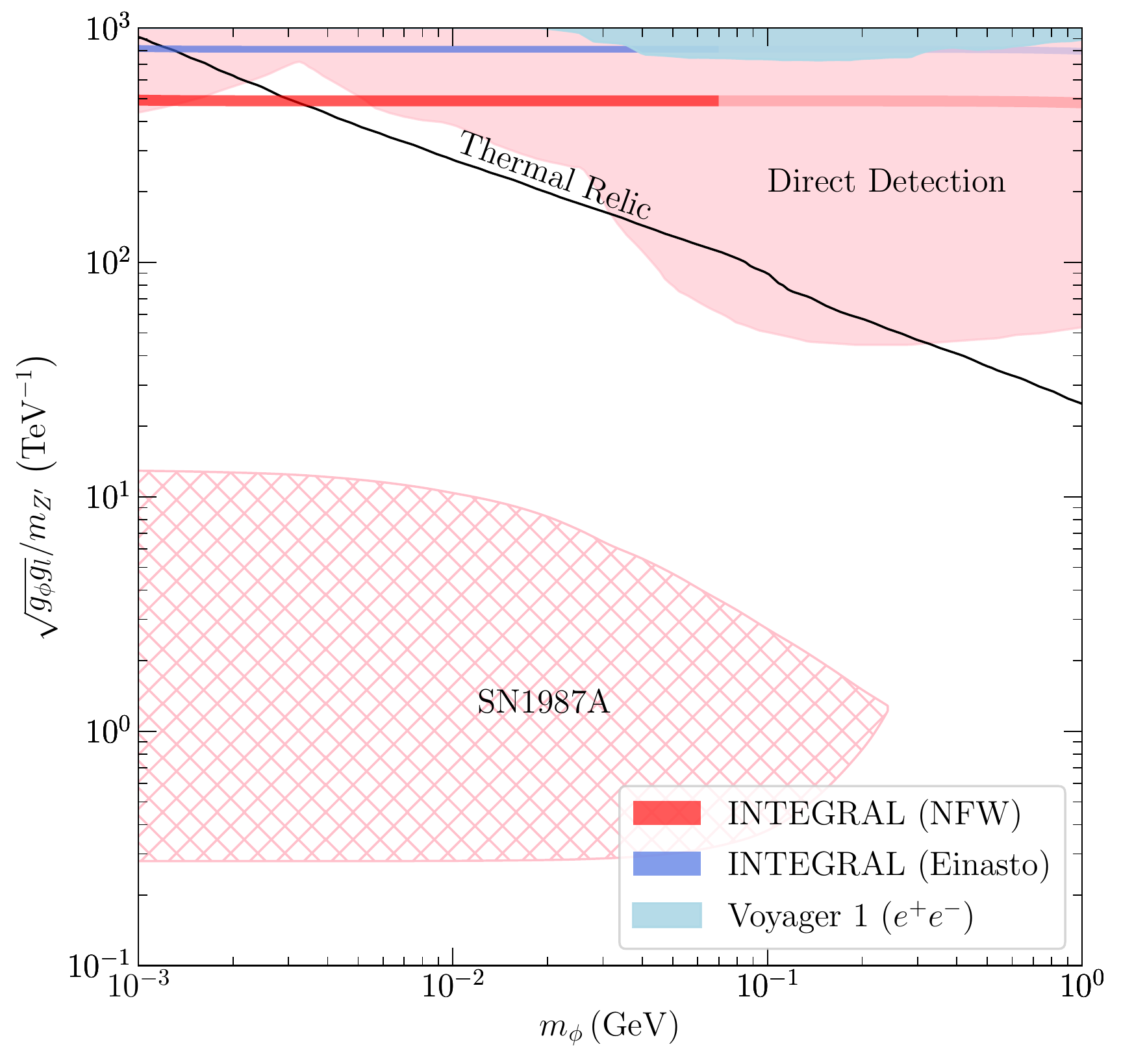}
    \caption{Bounds on the inverse of effective UV scale $\Lambda_{Z'}^{-1} = \sqrt{g_\phi g_l}/m_{Z'}$ for the $Z'$ model from laboratory tests (left panel) and from cosmological and astrophysical probes including direct detection (right panel).  The parameter regions of interest for the INTEGRAL excess are shown as thin blue and red bands; for $m_\phi \ge 70\,$MeV the DM interpretation is disfavored as indicated by a lighter shading.
    LEP bound only applies for $m_{Z'}$ above the EW scale, below which \eqref{eq:ZpLEP} applies instead. We do not show a band for $(g-2)_{\mu}$, which would need an assumption on $g_\phi/g_l$, since it is already excluded elsewhere (see main text and Fig.~\ref{Fig:summary_Z}).
     \label{Fig:constraint_Zp}}
\end{figure*}

\subsection{Direct detection}

As we focus on sub-GeV mass $\phi$ particles, we consider $\phi$-$e$ scattering signal as it has a lower threshold on DM mass in direct detection experiments.
Exclusion limits are customarily presented in terms of a reference scattering cross section~\cite{Essig:2011nj},
\begin{equation}
	 \bar{\sigma}_e = \dfrac{1}{16\pi (m_e + m_\phi)^2} \overline{|\mathcal{M}_{\phi \text{-} e} (q)|}^2_{q^2 = \alpha^2 m_e^2 }, 
\end{equation}
where $\overline{|\mathcal{M}_{\phi\text{-} e} (q)|}^2_{q^2 = \alpha^2 m_e^2 }$ is the squared matrix element of $\phi$ scattering on a free electron, summed over final state spins and averaged over initial state spin, evaluated at a typical atomic momentum transfer $q = \alpha m_e$.
To order $\mathcal{O}(v_{\rm rel}^2)$ it is given by
\begin{equation}
\begin{split}
	 &  \overline{|\mathcal{M}_{\phi\text{-} e} (q)|}^2_{F} = \dfrac{16 c_L^2 c_R^2 m_e^2}{m_F^2}, \\
	 & \overline{|\mathcal{M}_{\phi\text{-} e} (q)|}^2_{Z'} = \dfrac{16 g_\phi^2 g_l^2 m_\phi^2 m_e^2}{m_{Z'}^4} , 
\end{split}
\end{equation}
for the two representative models.
Note that  bounds on $\bar{\sigma}_e$ have been obtained for the present case of constant DM form factors, most recently in SENSEI~\cite{Barak:2020fql}, which also summarizes previous bounds from  XENON10~\cite{Essig:2012yx, Essig:2017kqs} and XENON1T~\cite{Aprile:2019xxb}, as well as from considering a solar-reflected DM flux~\cite{An:2017ojc}.
The corresponding constraints, combining the results of  experiments mentioned above,   are shown in Figs. \ref{Fig:constraint_F} and \ref{Fig:constraint_Zp}.

\subsection{Indirect search}

To explain the INTEGRAL signal, $\phi$ has to be a symmetric DM candidate, implying $\phi$ annihilation into SM leptons both at epochs of BBN and CMB, as well as at low redshift. Among them, bounds from BBN observables and DM annihilating to neutrinos~\cite{Hufnagel:2017dgo, Depta:2019lbe} are very weak, and are not further considered. 
Since in the considered models $\phi$ does not annihilate into photons at tree-level (except when accompanied by final state radiation), we focus on the channel $\phi \phi^* \rightarrow e^+ e^-$. 

For the $F$-mediated case, in which both $s$-wave and $p$-wave annihilation are present~\eqref{Eq:nonrel_annF_nsc}, it turns out that the constraint from CMB~\cite{Liu:2016cnk} is in general stronger than that from Voyager~1 data~\cite{Stone150}.
In the $Z'$-mediated case, since the leading contribution of $\phi \phi^* \rightarrow e^+ e^-$ is $p$-wave~\eqref{Eq:nonrel_annZ}, the annihilation at CMB epoch is velocity suppressed and the bounds from present-day data such as from Voyager~1~\cite{Boudaud:2016mos,Boudaud:2018oya} is more stringent, disfavoring DM masses above $\mathcal{O}(30)$\,MeV to explain the INTEGRAL 511\,keV line. This will be further improved by about one order of magnitude in future experiments, such as e-ASTROGAM\,\cite{DeAngelis:2016slk, Bartels:2017dpb} and AMEGO\,\cite{McEnery:2019tcm}. 
The CMB constraint for the $F$ case and the Voyager~1 constraint for the $Z'$ case  are shown in Figs.~\ref{Fig:constraint_F} and \ref{Fig:constraint_Zp}, respectively.

\subsection{Structure formation}

To avoid the collisional damping of DM primordial fluctuations~\cite{Boehm:2000gq,  Boehm:2004th}, DM has to kinetically decouple from the observable sector in the early Universe. In the considered models, DM couples to electrons and neutrinos with similar strength. 
Since the number density of electrons is much lower than that of background neutrinos once $T\ll m_e$, the scattering on neutrinos hence governs the ensuing constraint. 
 Here we take the bounds derived in \cite{Wilkinson:2014ksa, Arhrib:2015dez} for both energy-independent and energy-dependent DM-neutrino scattering cross sections.
Concretely, we require  that  for the $F$ model 
\begin{equation}
    \sigma_{\phi-\nu} \simeq {c_F^4 \over 8\pi m_F^2 } \,\,\lesssim  \,\, 10^{-36}\left({ m_\phi \over {\rm MeV} }\right)\,{\rm cm}^2\,,
\end{equation}
and for the $Z'$ model
\begin{equation}
    \sigma_{\phi-\nu} \simeq {E_\nu^2\, g_\phi ^2 g_l^2  \over 2\pi m_{Z'}^4 } \,\,\lesssim  \,\,10^{-41} \left({ m_\phi \over {\rm MeV} }\right)\left({ E_\nu \over    {\rm  eV} }\right)^2\,{\rm cm}^2\,. 
\end{equation}
The requirement consequently leads to   $c_F^2/m_F \lesssim 0.25\, (m_\phi/{\rm MeV})^{1/2}$\,TeV$^{-1}$, as well as $\sqrt{g_\phi  g_l}/m_{Z'} \lesssim 2.17\times 10^4\, (m_\phi/{\rm MeV})^{1/4}$\,TeV$^{-1}$.  Both bounds are weaker than those obtained above, and  are not shown in the figures.

\subsection{DM self-scattering}

If $\phi$ constitutes DM, its self-interaction may change the shape and density profile of DM halos, and the kinematics of colliding clusters. Such self-interaction is apparently very weak in the heavy $F$ model. 

In the $Z'$ model, the DM particle $\phi$ can efficiently self-scatter via  $Z'$ exchange.
The self-scattering cross section averaged over $\phi\phi\to \phi\phi$, $\phi\phi^*\to \phi\phi^*$ and $\phi^*\phi^*\to \phi^*\phi^*$
reads~\cite{Chu:2016pew}
\begin{equation}
   \sigma_{\rm SI}^{\phi\phi} ~ = ~ \dfrac{3g_\phi^4 m_\phi^2}{8\pi m_{Z'}^4}\,,
\end{equation}
where velocity-suppressed terms have been neglected.\footnote{At $m_{Z'} \sim 2m_\phi$  the velocity suppression  in $s$-channel $\phi\phi^*\to \phi\phi^*$ can be compensated by  the resonant enhancement. Such resonant contribution never dominates   and thus is not considered.}
However, the current bound, $\sigma_{\rm SI}/ m_\phi \leq 0.5\, {\rm cm}^2 /{\rm g}$ from cluster observations~\cite{Randall:2007ph,  Harvey:2015hha, Robertson:2016xjh,Harvey:2018uwf, Bondarenko:2017rfu}, is also not able to provide any meaningful bounds on the $Z'$ model with a $10\,$GeV $m_{Z'}$. 

\subsection{Anomalous Supernovae cooling}

An important constraint arises from the anomalous energy loss via $\phi$ production in hot stars, especially inside supernovae (SN), as we consider $m_\phi = \mathcal{O}({\rm MeV}\text{--}{\rm GeV})$ which has overlap with the SN core temperature. To avoid the suppression of neutrino emission from the SN core after explosion, we impose the so-called ``Raffelt criterion'', which states that the energy loss via dark particle production has to be smaller than  the luminosity in neutrinos, $L_{\nu}= 3\times 10^{52}$~erg/s~\cite{Raffelt:1996wa}.%
\footnote{The bounds from SN1987A are derived from the cooling of the proto-neutron star; doubts exist if SN1987A was a neutrino-driven explosion~\cite{Bar:2019ifz} in which case the limits become invalidated. Such speculation could be resolved once the remnant of SN1987A is firmly observed~\cite{Page:2020gsx}. } 
Here we follow the method in \cite{Chang:2018rso, Chu:2018qrm,Chu:2019rok}, and adopt the SN1987A numerical model of \cite{Fischer:2016cyd} with a total size $r_{\rm SN} = 35\,$km, to derive the bounds on the leptophilic DM models above. 

The dominant $\phi$ production channel is pair creation from  electron-positron annihilation. As our mediator particles, $F$ or $Z'$, are much heavier than the core temperature of SN, we can safely neglect thermal corrections.  Quantitatively, the lower boundaries of the exclusion regions are derived by requiring 
\begin{equation}
  \int^{r_{\rm c}}_0 d^3 r\int \frac{ d^3 p_{e^-} d^3 p_{e^+} }{ (2\pi)^6} f_{e^-}f_{e^+} \, (\sigma_{e^-  e^+ \to \phi\phi^*} v_M ) \sqrt{s} \lesssim L_{\nu}\,,
\end{equation} 
where $r_{\rm c}$ is the core size of SN1987A, taken as $15$\,km here and $f_{e^-, e^+}$ are Fermi-Dirac distributions for electron and positron.

On the flip side, if the coupling between $\phi$ and SM particles inside the SN core is so strong that the $\phi$ becomes trapped inside the  core, the energy loss via $\phi$ emission diminishes and again drops below the neutrino luminosity.\footnote{For even stronger couplings, the abundance  of $\phi$ particles trapped inside SN may help to capture SM neutrinos, leading to  an observable reduction in SN neutrino emission~\cite{Fayet:2006sa}. This may affect the parameter region studied in our Sec. \ref{Sec:onshellZp} for $m_\phi \lesssim 10\,$MeV. } To estimate the corresponding upper boundaries of couplings, we first define a radius $r_d$, where a
thermalized blackbody luminosity of $\phi$ equals $L_{\nu}$~\cite{Chang:2018rso, Chu:2018qrm}. Consequently, if $\phi$ gets further deflected by scattering off electrons for $r > r_d$, the effective dark luminosity drops below $L_{\nu}$. In practice, we impose
\begin{equation}
  \label{eq:sn1}
  \int^{r_{\rm SN}}_{r_d} dr \frac{1}{\lambda_{\phi \text{-} e}(r) }  \le 1 \,.
\end{equation}
Here, the  mean-free-path $\lambda_{\phi \text{-} e}$ at each radius is calculated by averaging the momentum distributions of both $\phi$ and electrons 
\begin{equation}
\lambda_{\phi \text{-} e} (r)= { \langle n_\phi v_\phi  \rangle  \over  \langle  n_e n_\phi   \sigma_{\phi \text{-} e}v_{M}    \rangle }\bigg|_r\,,%
\end{equation}
where the  Pauli blocking factor induced by final state electrons has been taken into account. The relevant differential cross sections are listed in App.~\ref{Sec:scattering}. 

The resulting SN1987A exclusion regions, combining both lower and upper boundaries, are given in the right panels of Figs.~\ref{Fig:constraint_F} and~\ref{Fig:constraint_Zp}. Our lower boundaries agree well with previous results~\cite{Guha:2018mli, Darme:2020ral}. Regarding the upper boundaries, we find that Pauli blocking plays an important role in suppressing $\phi$-electron scattering. Meanwhile, although there is little Pauli blocking in $\phi$-nucleon scattering, $\phi$ only couples to quarks at loop level, yielding a suppression by another factor $m_\phi^2/m_F^2$ for the $F$ model and $\alpha/\pi$ for the $Z'$ model.  It hence turns out that $\phi$-electron scattering dominates the capture in the parameter regions studied here.

\section{\boldmath$Z'$ mass below 10~GeV}
\label{Sec:onshellZp}

 \begin{figure*}[t]
    \centering
    \includegraphics[width=0.45\linewidth]{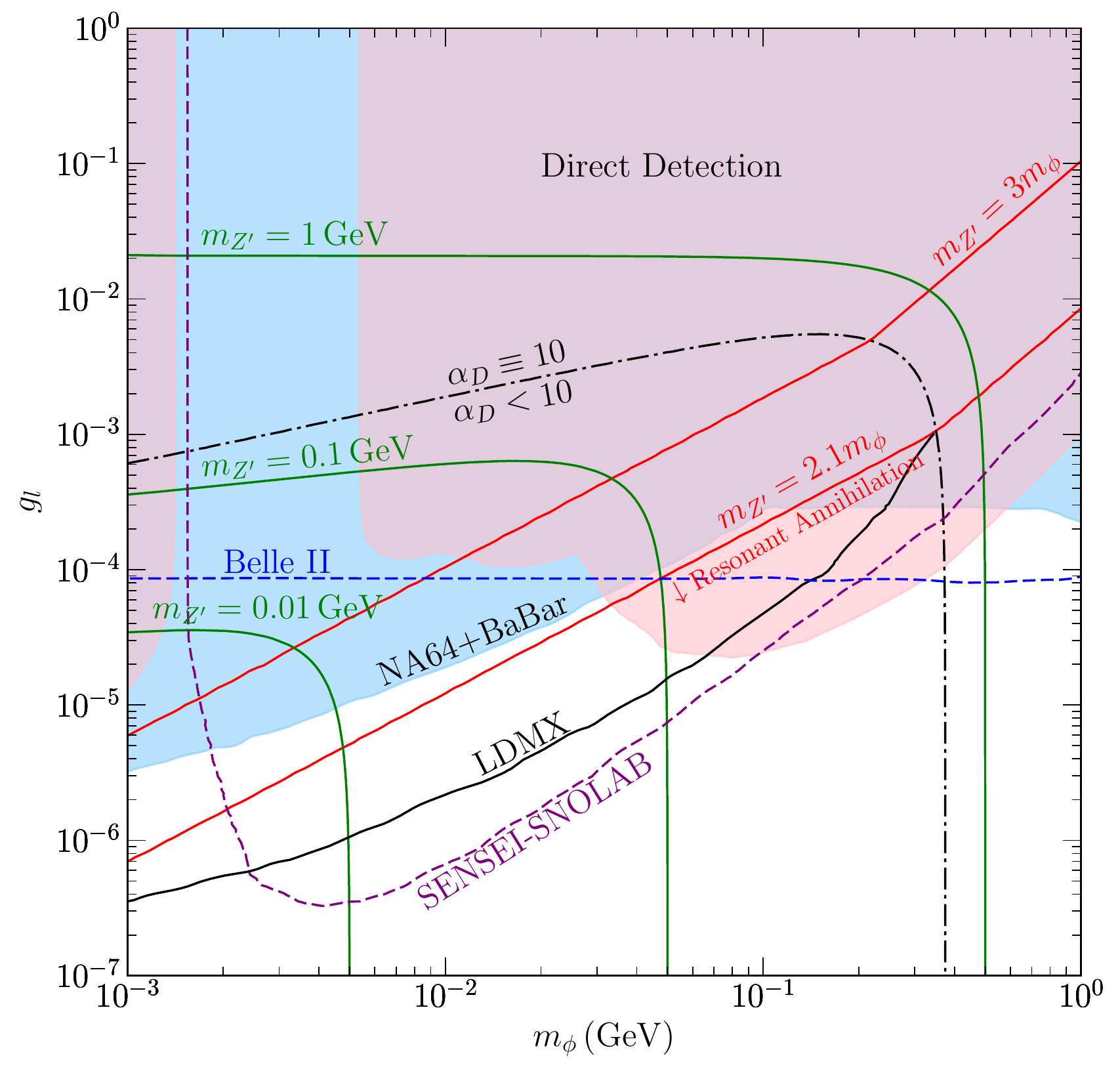}~~~~~
     \includegraphics[width=0.45\linewidth]{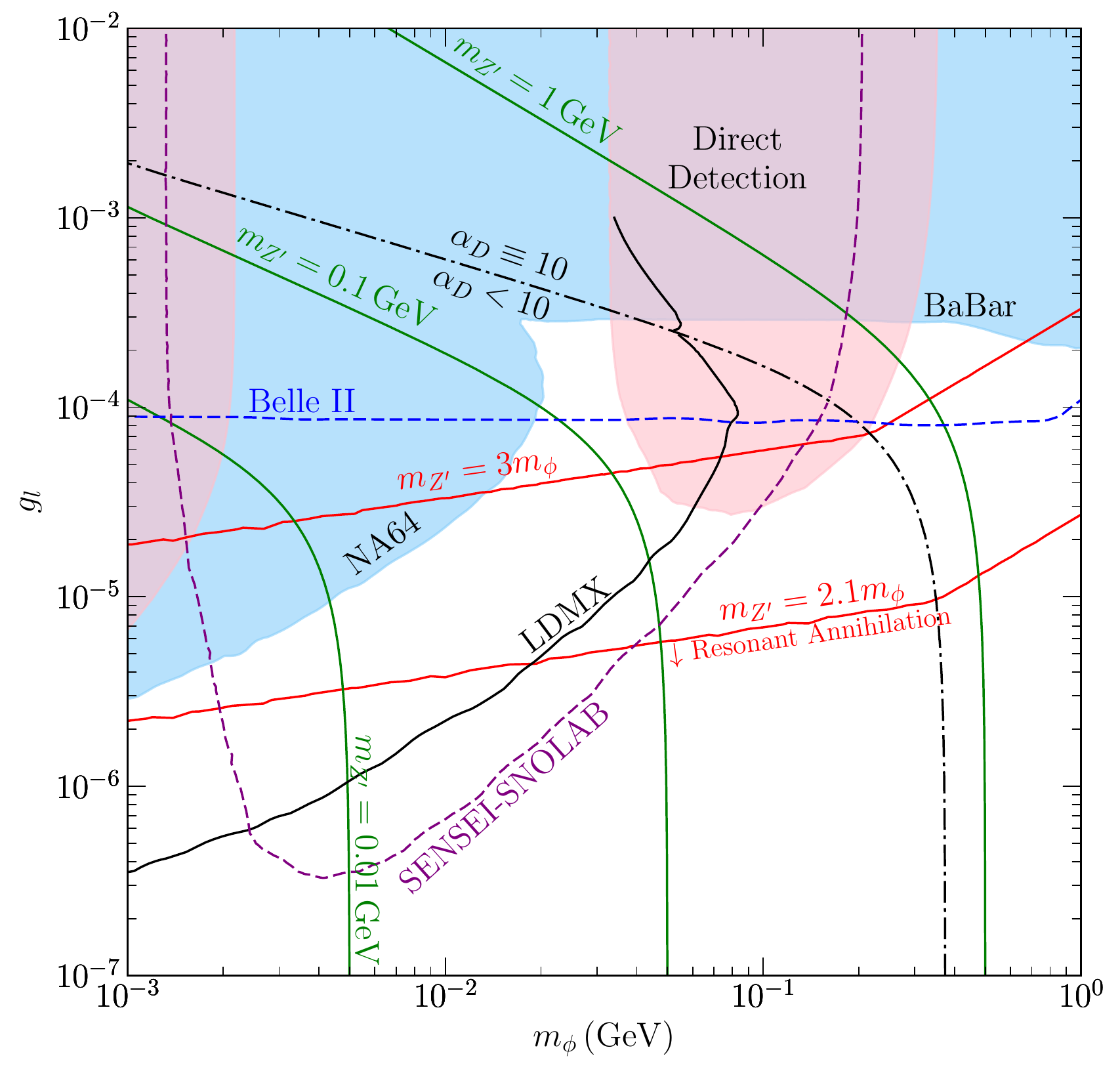}
    \caption{Conservative bounds on the lepton-$Z'$ coupling $g_l$ by saturating the cluster bound on the $\phi$-$Z'$ coupling $g_\phi$, and imposing either Eq.~\eqref{Eq:INTEGRAL_values} for INTEGRAL (left panel) or successful thermal freeze-out (right panel) to fix $m_{Z'}$. Below the  $m_{Z'} = 2.1m_\phi$ line, DM annihilation  happens resonantly at freeze-out or at low redshift. Additionally, Planck observations require $m_\phi\gtrsim 2$\,MeV in the minimal set-up. }
    \label{Fig:onshellZp}
\end{figure*}

Now we consider $m_{Z'} \lesssim 10$\,GeV, for which $Z'$ may appear on-shell in intensity-frontier experiments, CM energy permitting. In this case, the effective operator approach with its scale given by $\sqrt{g_\phi g_l }/ m_{Z'}$ does not apply any longer.  Nevertheless, the DM interpretation of the INTEGRAL signal  only relies on the product $g_\phi g_l$, instead of on each of the two couplings individually, up to a minor effect of $Z'$ decay width. %
Therefore, to obtain the most conservative constraints on $g_l$, we choose to maximize the dark coupling, $g_\phi$, using the cluster bounds~\cite{Harvey:2015hha,  Harvey:2018uwf, Bondarenko:2017rfu} above
\begin{equation} \label{SI:cluster}
    {\sigma_\text{SI}/ m_\phi}~\le ~ 0.5\,  {\rm cm^2/g}\,.
\end{equation}
We also impose a perturbativity bound, requiring that $\alpha_D = g_\phi^2/(4\pi)$ should not exceed~10.

There are then three parameters left. To explain the INTEGRAL signal,  Eq.~\eqref{Eq:INTEGRAL_values} allows to solve for the value of $m_{Z'}$ for each choice of ($m_\phi$, $g_l$). Now we directly adopt the existing constraints (e.g.~NA64 and BaBar) and projected sensitivities (e.g.~LDMX and Belle II) on an invisibly decaying $Z'$ from~\cite{Beacham:2019nyx}. For direct detection, we take the  bounds with electron recoils summarized in \cite{Barak:2020fql}, and future projections of SENSEI at SNOLAB in \cite{Emken:2019tni}. Interestingly, for $m_\phi$ below 100\,MeV, this may be further improved by orders of magnitude using potential signals from neutron star heating~\cite{Bell:2019pyc, Garani:2019fpa, Garani:2020wge}.  Indirect bounds on $p$-wave DM annihilation are relatively weaker~\cite{Depta:2019lbe}, and are not considered further.

The results are summarized  in the left panel of Fig.~\ref{Fig:onshellZp}.
Below the dash-dotted line, a self-interaction cross section saturating~\eqref{SI:cluster} is achieved with $\alpha_D< 10$. Above the line, we set $\alpha_D = 10$ and the bound~\eqref{SI:cluster} is not saturated.
 Assuming smaller values of $\alpha_D$  necessarily requires larger values of $g_l$ while their product has to be large enough to explain the INTEGRAL signal. Therefore, this choice of $\alpha_D$  gives the most conservative bounds, and choosing smaller $\alpha_D$ would exclude more parameter space in Fig.~\ref{Fig:onshellZp}. %
The green and red lines in the figure are the contours of $m_{Z'}$ and $m_{Z'}/m_{\phi}$, respectively. While  $m_{Z'}=3m_{\phi}$ has often been chosen in the literature, we also show the contour of $m_{Z'}=2.1m_{\phi}$, below which the average DM kinetic energy may overcome the mass barrier to produce intermediate $Z'$ on-shell at freeze-out or after.  As is shown in the figure, for each value of $m_{Z'}$ the mass ratio already becomes important in determining the annihilation cross section  at $m_{Z'} =  3m_{\phi}$. 

Our results suggest that most of the parameter region with non-resonant DM annihilation has been excluded, while the LDMX experiment is projected to probe the whole region of $m_{Z'} \gtrsim 2.1\,m_\phi$. This is different from the heavy $m_{Z'}$ case above, where Belle II always has a better sensitivity in probing the DM particle.  In presence of resonant annihilation, the required values of $g_l$ can be much smaller than what fixed-target experiments will reach, thus indirect DM searches, such as e-ASTROGAM\,\cite{Bartels:2017dpb}, are expected to better probe this parameter region. 

Similarly, instead of calibrating on INTEGRAL,  we may alternatively fix the annihilation cross section  by the thermal freeze-out requirement (right panel of Fig.~\ref{Fig:onshellZp}).   In this case, a limited parameter region is still allowed, but will be probed by future fixed-target experiments, as well as future direct detection experiments using electron recoils. Our result is in agreement with previous studies which generally fix $m_{Z'} \equiv  3m_\phi$, and $\alpha_D =0.1$~\cite{Beacham:2019nyx} (or $0.5$~\cite{Berlin:2018bsc, Akesson:2018vlm,Doria:2018sfx}),  but have not considered dark matter self-interaction. For both panels of Fig.~\ref{Fig:onshellZp}, the excluded region from $Z'$ searches would reduce by choosing larger values of $\alpha_D$.  By introducing the upper bound on $\alpha_D$ from Eq.~\eqref{SI:cluster}, the exclusion in Fig.~\ref{Fig:onshellZp} is hence robust in both cases, i.e.~either when imposing the DM interpretation of INTEGRAL (left)  or when requiring successful thermal freeze-out (right). 

\section{Conclusion}
\label{Sec:con}

In this work we consider the possibility that DM is a complex scalar particle $\phi$ with a mass below the GeV-scale. The particle is assumed to couple to SM leptons, either via a  charged fermion $F$ or via a vector boson $Z'$. These models fare among the simplest UV-complete extensions to the SM, and have been contemplated as sub-GeV DM candidates well before the field exploded with activity in this mass bracket. Among other reasons, they draw their attention from the fact that $\phi$ annihilation today might explain the galactic INTEGRAL excess and/or bring into reconciliation the prediction and observation of the anomalous magnetic moment of the muon.

Given the tremendous recent activity devoted to the search of light new physics, it is only timely to revisit these models of scalar DM in light of much new data. These particles can be probed in the laboratory such as in electron-beam experiments, and by astrophysical and cosmological observations. We collate the latest observational and experimental data and subject the model to all relevant bounds and provide forecasts on the sensitivity of proposed future experiments.

Respecting the bounds on charged particles from high energy colliders LEP and LHC, we consider $F$ to be at or above the EW scale. The combination $m_F/c_F^2$ is inherent to most observables and can be interpreted as the effective UV-scale $\Lambda_F$ for that model. We calculate the production of $\phi$-pairs, mediated by the exchange of off-shell $F$, in the fixed target experiments NA64, mQ, LDMX, and BDX as well in $e^+e^-$ colliders BaBar and Belle II. When the production is kinematically unsuppressed, the best bound is $\Lambda_F \gtrsim 250\,\GeV$ by BaBar, currently surpassed by LEP with  $\Lambda_F  \gtrsim 1\,\TeV$. LDMX-II can improve on this number to $5~\TeV$. Turning to the  $Z'$ model, we  consider both heavy and light vector mediators. If $Z'$ remains off-shell in all experiments, we may take the combination $m_{Z'}/\sqrt{g_\phi g_l}$ as the effective UV scale $\Lambda_{Z'}$. In this case, BaBar points to $\Lambda_{Z'} \gtrsim 35\,\GeV$ to be improved by Belle-II to  $\Lambda_{Z'} \gtrsim 170\,\GeV$ at best, weaker than the current LEP bound of $\Lambda_{Z'} \gtrsim 346\,\GeV$. 

These direct limits are then compared to loop-induced precision observables, concretely, to $g-2$, to the invisible width of the $Z$ and to $Z$-boson oblique corrections. We explicitly revisit all those calculations, confirming  previously presented scaling relations in the limit $m_{\phi,l}/m_F \ll 1$ or $m_{\phi,l}/m_{Z'}\ll 1$, and, as an added value, provide the full expressions of the loop integrals. We find that for the $F$  model, the improved limit obtained from $g-2$ of the electron surpasses all direct observables, with $\Lambda_F\gtrsim 10^4\,\TeV$, while for the $Z'$ model, they do not play a role in the phenomenology. We also complement those constraints with limits that arise from the freedom in the chiral structure of the models, using the parity asymmetry in polarized electron scattering. Finally, we derive limits from lepton flavor violation that are dependent on the concrete UV-content of the models.

Turning to astrophysical constraints, we derive the anomalous energy loss induced by $\phi$-pair production in the assumed proto-neutron star of SN1987A. This adds strong and complementary new limits on the parameter space for $m_\phi\lesssim 100\,\MeV$ down to $\Lambda_F \gtrsim 10^5\,\TeV$ and $\Lambda_{Z'}\gtrsim 3\,\TeV$. We furthermore consider constraints from direct detection, structure formation, CMB energy injection, and DM-self scattering. Here, the CMB puts stringent constraints on the $s$-wave annihilation mediated by $F$. In turn, for the $p$-wave annihilation mediated by $Z'$ the bounds are sub-leading. For those reasons, a thermal freeze-out in the $F$-mediated model is  firmly excluded, whereas the $Z'$ model remains little constrained from  energy injection.

Regarding the DM interpretation of the INTEGRAL 511\,keV line, we show that it is excluded in both the $F$-mediated model as well as in the $Z'$-mediated model with $m_{Z'} \ge 10\,$GeV. In the model with charged $F$, the crucial constraints come from the $(g-2)_e$ data, from the CMB,   and from SN1987A. For the $Z'$ model, intensity-frontier experiments and direct detection via electron recoils play the major role. However, a caveat exists: if the annihilation is resonant, $m_{Z'} \simeq 2 m_\phi$,  the INTEGRAL signal may still be explained in conjunction with a light $m_{Z'} \le 10\,$GeV while at the same time being experimentally allowed.%

In a final part, we then entertain the possibility of a $Z'$-mass below 10\,GeV. The phenomenology then changes, as the vector may go on-shell in  various considered searches. To derive conservative constraints on the $Z'$ coupling to the observable sector, $g_l$, we then saturate the dark coupling $g_\phi$, taking its maximally allowed value from DM self-scattering.
Requiring a successful explanation of INTEGRAL by fixing $m_{Z'}$ constrains the values of the remaining free parameters $g_l$ and $m_\phi$, driving $m_\phi$ to the resonant annihilation regime; the combination of LDMX, Belle II, and direct detection via electron recoils will be able to completely cover the parameter region of $m_{Z'}\ge  2.1\,m_\phi$ in the near future, \text{i.e.}~the range where $Z'$ remains off-shell.

As an outlook, we comment on the resonant region which is not studied here. For $ m_{Z'} \simeq  \,2m_\phi$ the annihilation cross section is greatly enhanced and the required value on $g_l$ coming from the annihilation cross section diminishes. This hampers the direct experimental sensitivity considered in this work. In turn, however, it opens the possibility of using displaced vertex searches in fixed-target experiments, depending on the decay mode of $Z'$. 
Dialing down the $Z'$-mass further, $m_{Z'} < 2 m_\phi $ the annihilation via $\phi\phi^*\to Z'^{(*)} Z'^{(*)} \to 2e^+ 2e^-$ will eventually come to dominate.  As the process is not velocity suppressed, we then re-enter the regime where stringent  CMB bounds apply. We leave those aspects to dedicated future work.

\vspace{.3cm}

\paragraph*{Acknowledgments}
 XC and JP are supported by the New Frontiers program of the Austrian Academy of Sciences. JLK is supported by the Austrian Science Fund FWF under the Doctoral Program W1252-N27 Particles and
  Interactions. We thank the Erwin Schr\"odinger International Institute for Mathematics and Physics where this work was started for their hospitality. We acknowledge the use of computer packages for
  algebraic calculations~\cite{Mertig:1990an,Shtabovenko:2016sxi}.

\appendix

\section{Full expression for \boldmath$(g-2)_l$}
\label{app:g-2}

From the explicit calculations for the $(g-2)_l$ contribution in the representative models as shown in Fig.~\ref{Fig:gminus2}, we obtain,
\begin{align}
     &\Delta a_l^{(F)} = \int_0^1 dz\, \dfrac{m_l (z-1)^2 \left[ m_l z (c_L^{l\,2} + c_R^{l\,2}) + 2 c_L^l c_R^l m_F \right]}{16\pi^2 \big[ m_F^2 (1-z) + m_\phi^2 z  \big]}\,,\nonumber \\
    &\Delta a_l^{(Z')} = \int_0^1 dz\, \dfrac{m_l^2 z (z-1)\left[ g_L^2 (z+1) - 4 g_L g_R + g_R^2 (z+1)\right]}{8\pi^2 \big[ m_l^2 (z-1)^2 + m_{Z'}^2 z \big]}\,.
\end{align}
Taking only the leading order terms and setting $g_L = g_R \equiv g_l$, we recover Eq.~\eqref{Eq:g-2_F} and Eq.~\eqref{Eq:g-2_Zp} of the main text.

\section{Formulae for cross sections}

\subsection{\boldmath$\phi$-pair annihilation}
\label{Sec:phi_pair_ann}

Here we give the non-relativistic expansion of the DM annihilation cross section via a $t$-channel fermion $F$, to second order of the relative velocity, assuming the hierarchy $m_F \gg m_\phi >m_l$,
\begin{widetext}
{
\medmuskip=0mu
\thinmuskip=0mu
\thickmuskip=0mu
\nulldelimiterspace=1pt
\scriptspace=0pt
\begin{eqnarray}
\label{eq:annFlong}
    &\sigma_{{\rm ann},F} v_M  =  \dfrac{( c_L^2 m_l + c_R^2 m_l   +2 c_L c_R m_F)^2 }{16 \pi (m_F^2+m_\phi^2 - m_l^2)^2} \left(1 - \dfrac{m_l^2}{m_\phi^2} \right)^{\tfrac{3}{2}} %
    + v_{\rm rel}^2 \left[ \dfrac{(c_L^4-12 c_L^2 c_R^2+c_R^4)m_\phi^2 }{48 \pi  m_F^4} + \dfrac{3c_L^2 c_R^2 m_l^2 }{32\pi m_F^2 m_\phi^2} 
    -  \dfrac{ (c_L^4-14c_L^2 c_R^2 +c_R^4) m_l^2 }{64\pi m_F^4} %
    \right].
\end{eqnarray}
}
\end{widetext}
Here we have neglected terms of $ 
 {O}({m_{l,\phi}^5}/{m_F^5})$.
By taking only the leading order in ${O}(m_{l,\phi}/m_F)$ and replacing $v_{\rm rel}$ with $2 v_\phi$, we retrieve the $s$-wave expression of Eq.~(1) in~\cite{Boehm:2003hm}, while our $p$-wave result differs. 
The difference arises from the general mismatch between $v_{\rm rel}$ and $v_M$. 
We prefer to use the non-relativistic expansion of the Lorentz invariant quantity $\sigma_{\rm ann} v_M$~\cite{Gondolo:1990dk}; our scattering amplitude agrees with the one given in the appendix of~\cite{Boehm:2003hm}.
For the special case $c_L c_R =0$ and for $m_l =0$, the $s$-wave component in~\eqref{eq:annFlong} vanishes and the process becomes $p$-wave dominated, scaling as $v_{\rm rel}^2  m_\phi^2/m_F^4$. 

For real scalar DM, $\phi = \phi^*$, the annihilation process via a $u$-channel fermion $F$   needs to be added, and the cross section in the same approximation becomes
\begin{widetext}
{
\medmuskip=0mu
\thinmuskip=0mu
\thickmuskip=0mu
\nulldelimiterspace=1pt
\scriptspace=0pt
\begin{eqnarray}
\label{eq:B2}
    & \sigma^{\text{real }\phi}_{{\rm ann},F} v_M  =  \dfrac{(c_L^2 m_l  +  c_R^2 m_l  +2 c_L c_R m_F)^2 }{4 \pi (m_F^2+m_\phi^2 - m_l^2)^2} \left(1 - \dfrac{m_l^2}{m_\phi^2} \right)^{\tfrac{3}{2}}  + v_{\rm rel}^2 \Bigg[   \dfrac{ 3c_L^2 c_R^2  m_l^2 }{8\pi m_F^2 m_\phi^2 } + \dfrac{ 3c_Lc_R(c_L^2+c_R^2) m_l^3}{ 8\pi m_F^3m_\phi^2}  - \dfrac{ c_L^2 c_R^2 m_\phi^2 }{ \pi m_F^4}    +  \dfrac{ 3c_L^2 c_R^2 m_l^2 }{4\pi m_F^4} \Bigg].
\end{eqnarray}
}
\end{widetext}

\subsection{\boldmath$e^- e^+$ annihilation with initial state radiation}
\label{Sec:epem_collider_xsec}
 First, we detail the annihilation cross section associated with initial state radiation (ISR), corresponding to Fig.~\ref{Fig:ecollider}.
 Following~\cite{Chu:2018qrm}, the differential cross section with ISR is formulated as the cross section without ISR times the improved Altarelli-Parisi radiator function. 
The annihilation cross sections, without ISR and with $s$ denoting the squared CM energy,  to order ${O}({m_{e,\phi}^3}/{m_F^3})$ and after an average over the initial state spins has been performed, reads
\begin{align}
     \sigma_{e^- e^+ \rightarrow \phi \phi }^{(F)} &=   \dfrac{c_L^2 c_R^2 }{32 \pi m_F^2}  \sqrt{\dfrac{s-4m_\phi^2}{s-4m_e^2}}\left(1- \dfrac{4m_e^2}{s}\right) , \\
     \sigma_{e^- e^+ \rightarrow \phi \phi }^{(Z')}  &= g_\phi^2 (s - 4m_\phi^2) \sqrt{\dfrac{s - 4m_\phi^2}{s-4m_e^2}}  \notag \\
  	&\quad \times \dfrac{s (g_L^2 +g_R^2 ) - m_e^2 (g_L^2 - 6 g_L g_R + g_R^2)}{96\pi s (s - m_{Z'}^2)^2} .
\end{align}

\subsection{\boldmath$e$-$N$ bremsstrahlung}
\label{Sec:eNbrem}

For the fixed-target experiments, we consider the production of $\phi$-pairs via electron-nucleus bremsstrahlung depicted in Fig.~\ref{Fig:fixed-target}.
Here, we need to compute the 2-to-4  cross section $\sigma_{2\rightarrow 4}$ for the process $e^- (p_1) + N(p_2) \rightarrow e^- (p_3) + X_n (p_4) + \phi (p_\phi) + \phi^* (p_{\phi^*})$.
We define $q = p_\phi + p_\phi^*$ with $q^2 \equiv s_{\phi\phi} $, $q_1 = p_1 - p_3$ with $q_1^2 \equiv t_1 $ and $q_2 = p_2 - p_4$ with $ q_2^2 \equiv t_2$ such that $q = q_1 + q_2$ given that total momentum is conserved.
The differential cross section is then written as 
\begin{equation}
\label{Eq:diff2to4}
       d\sigma_{2\rightarrow 4} = \dfrac{1}{4 E_1 E_2 v_M } \dfrac{1}{g_1 g_2} \sum\limits_{\rm spins} |\mathcal{M}|^2 d \Phi\,,	
\end{equation}
where $g_1$ and $g_2$ are spin degrees of freedom of electron and nucleus, respectively, while $|\mathcal{M}|^2$ is the amplitude square, and $d\Phi$ is the total phase space. 
By introducing an integral w.r.t. $s_X \equiv m_X^2 = p_4^2$, in the lab frame Eq.~\eqref{Eq:diff2to4} becomes
\begin{equation}
   d\sigma_{2\rightarrow 4}=  \dfrac{(4\pi \alpha)^2}{2| \vec{p}_1 | m_N g_2 q_2^4 } L^{\mu\nu, \rho \sigma} \phi_{\rho \sigma} W_{\mu\nu} (-q_2) ds_X d\Phi_4\,,\nonumber
\end{equation}
where $L^{\mu\nu, \rho \sigma}$ contains the leptonic average over $g_1$, $\phi_{\rho \sigma}$ includes the $\phi$-emission piece together with the heavy mediator propagator, $W_{\mu\nu}$ is the hadronic tensor with its concrete form given in~\cite{Chu:2018qrm} and $d\Phi_4$ is the 4-body phase space which is also analytically computed in~\cite{Chu:2018qrm}. 

The leptonic tensor from the two diagrams and their interference can be expressed as
\begin{align}
L^{\mu\nu, \rho\sigma} &=  \dfrac{L^{\mu\nu, \rho\sigma}_{a} }{\big[ (p_3 + q)^2 -m_e^2\big]^2} + \dfrac{L^{\mu\nu, \rho\sigma}_{b}}{\big[(p_1 - q)^2 -m_e^2	\big]^2} \nonumber \\&\quad + \dfrac{L^{\mu\nu, \rho\sigma}_{ab}}{\big[(p_3 + q)^2 -m_e^2\big] \big[(p_1 - q)^2 -m_e^2	\big]}\,,
\end{align}
where, for completeness, we spell out the individual traces in the following. 
For the $F$-mediated model they read,
\begin{widetext}
\begin{align}
	 & L^{\mu\nu, \rho\sigma}_{a, F} = \dfrac{1}{g_1 m_F^2} {\rm Tr} \Big [(\slashed{p}_3 + m_e) (c_L P_L +c_R P_R)(c_L P_R + c_R P_L)(\slashed{p}_3 + \slashed{q} + m_e)\gamma^\mu (\slashed{p}_1 + m_e)\gamma^\nu  \notag \\
	 &  \quad\quad\quad\quad\quad  (\slashed{p}_3 + \slashed{q} + m_e)(c_L P_L +c_R P_R)(c_L P_R + c_R P_L) \Big],   \notag\\
	 &  L^{\mu\nu, \rho\sigma}_{b, F} = \dfrac{1}{g_1 m_F^2} {\rm Tr} \Big [(\slashed{p}_1 + m_e)(c_L P_L + c_R P_R) (c_L P_R +c_R P_L)(\slashed{p}_1 - \slashed{q} + m_e)\gamma^\nu (\slashed{p}_3 + m_e)\gamma^\mu    \notag \\
	 & \quad\quad\quad\quad\quad  (\slashed{p}_1 - \slashed{q} + m_e) (c_L P_L +c_R P_R)(c_L P_R +c_R P_L) \Big],   \notag\\
	 &  L^{\mu\nu, \rho\sigma}_{ab, F} = \dfrac{1}{g_1 m_F^2} \bigg\lbrace {\rm Tr} \Big[ (\slashed{p}_3 + m_e)(c_L P_L +c_R P_R) (c_L P_R + c_R P_L)(\slashed{p}_3 + \slashed{q} + m_e) \gamma^\mu (\slashed{p}_1 + m_e) (c_L P_L +c_R P_R)    \notag \\
	  &\quad\quad\quad\quad\quad   (c_L P_R +c_R P_L)(\slashed{p}_1 - \slashed{q} + m_e)\gamma^\nu  \Big] +  {\rm Tr} \Big[(\slashed{p}_3 + m_e) \gamma^\mu  (\slashed{p}_1 - \slashed{q} + m_e) (c_L P_L +c_R P_R) (c_L P_R +c_R P_L)    \notag\\
	  & \quad\quad\quad\quad\quad  (\slashed{p}_1 + m_e) \gamma^\nu (\slashed{p}_3 + \slashed{q} + m_e) (c_L P_L +c_R P_R)  (c_L P_R +c_R P_L)  \Big] \bigg\rbrace\,, \label{eq:appL}
\end{align}
where indices $\rho, \sigma$ can obviously be abandoned. For the vector-mediated model we obtain for the traces,
\begin{align}
	 & L^{\mu\nu, \rho\sigma}_{a, Z'} = \dfrac{1}{g_1} {\rm Tr} \Big[ (\slashed{p}_3 + m_e)\gamma^\rho (g_L P_L +g_R P_R)(\slashed{p}_3 + \slashed{q} + m_e)\gamma^\mu (\slashed{p}_1 + m_e)\gamma^\nu (\slashed{p}_3 + \slashed{q} + m_e) (g_L P_R +g_R P_L)\gamma^\sigma \Big], \notag \\
	 &  L^{\mu\nu, \rho\sigma}_{b, Z'} = \dfrac{1}{g_1} {\rm Tr} \Big[(\slashed{p}_1 + m_e) (g_L P_R +g_R P_L)\gamma^\sigma(\slashed{p}_1 - \slashed{q} + m_e)\gamma^\nu (\slashed{p}_3 + m_e)\gamma^\mu (\slashed{p}_1 - \slashed{q} + m_e)\gamma^\rho (g_L P_L +g_R P_R) \Big],  \notag\\
	 &  L^{\mu\nu, \rho\sigma}_{ab, Z'} = \dfrac{1}{g_1} \bigg\lbrace {\rm Tr} \Big[(\slashed{p}_3 + m_e)\gamma^\rho (g_L P_L +g_R P_R)(\slashed{p}_3 + \slashed{q} + m_e)\gamma^\mu (\slashed{p}_1 + m_e) (g_L P_R +g_R P_L)\gamma^\sigma (\slashed{p}_1 - \slashed{q} + m_e) \gamma^\nu \Big] \notag\\
	  &\quad\quad\quad\,\,\,   +  {\rm Tr} \Big[ (\slashed{p}_3 + m_e)\gamma^\mu (\slashed{p}_1 - \slashed{q} + m_e)\gamma^\rho(g_L P_L +g_R P_R) (\slashed{p}_1 + m_e)\gamma^\nu (\slashed{p}_3 + \slashed{q} + m_e) (g_L P_R +g_R P_L)  \gamma^\sigma \Big]  \bigg\rbrace .
\end{align}
\end{widetext}

In the case that  $\phi$-emission proceeds by $F$ mediation, we may simplify the calculation by utilizing the fact that $m_F = 1\,{\rm TeV}$ is much larger than any momentum transfer considered here and approximate the the $F$ propagator as 
\begin{equation}
     \dfrac{i \left(\slashed{q} + m_F \right)}{q^2 - m_F^2} \rightarrow \dfrac{-i}{m_F}\,,
\end{equation}
where $q$ is the four-momentum flowing in the propagator.
We hence write the $l l \phi\phi$ interaction as an effective operator suppressed by $m_F$.
The $\phi$-emission piece is thus $\phi_{\rho\sigma}^{(F)} = 1$, and the integral over the product of Lorentz-invariant phase space of $\phi$ and $\phi^*$ denoted by $d\Pi_{\phi, \phi^*}$, is given by
\begin{equation}
	\int d\Pi_{\phi, \phi^*} (2\pi)^4 \delta^{(4)} ( q- p_\phi -p_{\phi^*} ) \phi_{\rho\sigma}^{(F)} = \dfrac{1}{8\pi} 	\sqrt{1- \dfrac{4 m_\phi^2}{s_{\phi\phi}} }\,.\nonumber
\end{equation}
Note that the Lorentz indices $\rho,\,\sigma$ do not have physical meaning in the $F$-mediated model, and are introduced to unify the notation of the two models.

The $\phi$-emission piece $\phi_{\rho \sigma}^{(Z')}$ in the $Z'$ case reads
\begin{align}
\phi_{\rho \sigma}^{(Z')} &= \dfrac{g_\phi^2}{\left( s_{\phi\phi} - m_{Z'}^2 \right)^2 } \left( q^\alpha q^\beta - 2 p_\phi^\alpha p_{\phi^*}^\beta - 2 p_{\phi^*}^\alpha p_{\phi}^\beta \right) \notag  \\
&\quad \times \left( g_{\rho\alpha} -  \dfrac{q_\rho q_\alpha}{m_{Z'}^2}\right) \left( g_{\sigma\beta} -  \dfrac{q_\sigma q_\beta}{m_{Z'}^2}\right).
\end{align}
Terms containing $p_\phi$ and $p_{\phi^*}$ can be integrated over the $\phi$-pair phase space using a modified version of Lenard's formula~\cite{Chu:2019rok} for massive final states, yielding 
\begin{align}
&\int d\Pi_{\phi, \phi^*} (2\pi)^4 \delta^{(4)} ( q- p_\phi -p_{\phi^*} ) \phi_{\rho\sigma}^{(Z')} \notag \\
&= \dfrac{1}{8\pi  \left(s_{\phi\phi} - m_{Z'}^2 \right)^2 } 	\sqrt{1- \dfrac{4 m_\phi^2}{s_{\phi\phi}} }  f_{Z'}(s_{\phi\phi} ) \left(-g_{\rho\sigma} + \dfrac{q_\rho q_\sigma}{s_{\phi\phi}} \right).\nonumber
\end{align}
Here, the function $f(s_{\phi\phi})$  following the convention used in~\cite{Chu:2018qrm,Chu:2019rok,Chu:2020ysb} and for the $Z'$-mediated model reads 
\begin{equation}
 f_{Z'}(s_{\phi\phi})  = \dfrac{1}{3} g_\phi^2  s_{\phi\phi}  \left( 1 - \dfrac{4 m_\phi^2}{s_{\phi\phi}}  \right).
\end{equation}

When considering the experiments NA64 and LDMX where final state electrons are measured, we  utilize the following differential cross section %
\begin{widetext}
\begin{align}
	&\dfrac{d\sigma_{2\rightarrow 4}^{F}}{dE_3 d\cos\theta_3} =  \dfrac{\alpha^2}{8\pi^2  g_2m_N} \sqrt{\dfrac{E_3^2 -m_e^2}{ E_1^2 -m_e^2}}\int ds_X \int ds_{\phi\phi} \int d t_2 \int dp_{1q} \left| \dfrac{\partial \phi_4^{R4q}}{\partial p_{1q}} \right|  \dfrac{1}{t_2^2}\dfrac{1}{\sqrt{\lambda (s_4, m_N^2, t_1)}} \notag \\
	&\quad\quad\quad\quad\quad\, \times L^{\mu\nu, \rho \sigma}_{F} W_{\mu\nu} (-q_2) \dfrac{1}{16\pi^2  } 	\sqrt{1- \dfrac{4 m_\phi^2}{s_{\phi\phi}} }  , \\
	&\dfrac{d\sigma_{2\rightarrow 4}^{Z'}}{dE_3 d\cos\theta_3} =  \dfrac{\alpha^2}{8\pi^2 g_2 m_N} \sqrt{\dfrac{E_3^2 -m_e^2}{ E_1^2 -m_e^2}}\int ds_X \int ds_{\phi\phi} \int d t_2 \int dp_{1q} \left| \dfrac{\partial \phi_4^{R4q}}{\partial p_{1q}} \right|  \dfrac{1}{t_2^2}\dfrac{1}{\sqrt{\lambda (s_4, m_N^2, t_1)}} \notag \\
	&\quad\quad\quad\quad\quad\, \times L^{\mu\nu, \rho \sigma}_{Z'} W_{\mu\nu} (-q_2) \dfrac{1}{16\pi^2  \left(s_{\phi\phi} - m_{Z'}^2 \right)^2 } 	\sqrt{1- \dfrac{4 m_\phi^2}{s_{\phi\phi}} }  f_{Z'}(s_{\phi\phi} ) \left(-g_{\rho\sigma} + \dfrac{q_\rho q_\sigma}{s_{\phi\phi}} \right)\,,  
\end{align}
\end{widetext}
where $p_{1q} \equiv p_1 \cdot q$, the angle $\phi_4^{R4q}$ is the angle between $(\vec{q}_1, \vec{p}_1)$ plane and $(\vec{q}_1, \vec{q})$ plane in the frame that $\vec{p}_4 + \vec{q} = 0$, $s_4 \equiv (p_4 + q)^2 = (p_1 + p_2 -p_3)^2$, $\lambda(a,b,c) = a^2 +b^2 +c^2 -2 ab -2bc -2ac$ is the K\"{a}ll\'{e}n function, and $E_1 = E_{\rm beam}$ in the lab frame. 
The integration boundaries for $s_{\phi\phi}$, $t_2$ and $p_{1q}$, and the Jacobian $|\partial \phi_4^{R4q} / \partial p_{1q}|$ are found in~\cite{Chu:2018qrm,Chu:2019rok}.
In the lab frame ($|\vec{p}_2| =0$), the Lorentz-invariant variables $s_4$ and $t_1$ can be expressed in terms of $E_1$, $E_3$ and $\cos \theta_3$ as 
\begin{align}
	  s_4 &= 2m_e^2 +m_N^2 +2E_1 m_N - 2E_3 m_N - 2E_1 E_3 \notag \\&\quad + 2 \sqrt{E_1^2 -m_e^2}  \sqrt{E_3^2 -m_e^2}  \cos\theta_3\,, \nonumber\\
	  t_1 &= 2m_e^2 -2E_1 E_3 + \sqrt{E_1^2 -m_e^2}  \sqrt{E_3^2 -m_e^2}  \cos\theta_3\,.\nonumber
\end{align}

For the experiments mQ and BDX  we need the spectrum and distribution of the produced $\phi$ particles. Therefore, we use
\begin{widetext}
\begin{align}
	\dfrac{d\sigma_{2\rightarrow 4}^{F,Z'}}{dE_\phi d\cos\theta_\phi} & =  \dfrac{\alpha^2}{16 (2\pi)^5 g_2 m_N} \sqrt{\dfrac{E_\phi^2 -m_\phi^2}{ E_1^2 -m_e^2}}\int ds_X \int ds_{36} \int d t_2 \int d\phi_{436}^* \int t_{23} \int d \phi_{36}^*  \nonumber \\
	& \times \dfrac{1}{t_2^2} \dfrac{1}{\sqrt{\lambda (s_{436}, m_N^2, t_{15})}} \dfrac{1}{\sqrt{\lambda (s_{36}, m_N^2, (p_1 - p_4 -p_\phi)^2 )}}  L^{\mu\nu, \rho \sigma}_{F, Z'} W_{\mu\nu} (-q_2) \phi_{\rho\sigma}^{(F, Z')},
\end{align}
\end{widetext}
where $s_{36} \equiv (p_3 + p_{\phi^*} )^2$, $\phi_{436}^*$ is the angle between the planes of $(\vec{p}_1 - \vec{p}_\phi , \vec{p}_1 )$ and  $(\vec{p}_1 - \vec{p}_\phi , \vec{p}_4 )$ in the frame that $\vec{p}_3 + \vec{p}_4 + \vec{p}_{\phi^*} = 0 $ ranging from $0$ to $2\pi$, $t_{23} \equiv (p_2 - p_3)^2$, $\phi_{36}^*$ is the angle between the planes of $(\vec{p}_1 - \vec{p}_4 - \vec{p}_\phi , \vec{p}_4 )$ and  $(\vec{p}_1 -\vec{p}_4 - \vec{p}_\phi , \vec{p}_3 )$ in the frame that $\vec{p}_3 + \vec{p}_{\phi^*} = 0 $  ranging from $0$ to $2\pi$, $s_{436} \equiv (p_3 +p_4 + p_{\phi^*})^2 = (p_1 +p_2 -p_\phi)^2$, and $t_{15}\equiv (p_1 -p_\phi)^2$.
The integration boundaries of these variables are given in~\cite{Chu:2018qrm}.
Note that 
\begin{equation}
(p_1 - p_4 -p_\phi)^2 = m_N^2 + s_X +s_{36} +t_{15} -s_{436} -t_2\,.\nonumber
\end{equation}
In the lab frame, one obtains the relations 
\begin{align}
	   s_{436} &=  m_e^2 +m_N^2 +m_\phi^2 +2E_1 m_N -2 E_\phi m_N - 2E_1 E_\phi \nonumber\\&\quad +2 \sqrt{E_1^2 -m_e^2 } \sqrt{E_\phi^2 -m_\phi^2} \cos\theta_\phi \,,\nonumber \\
	   t_{15} &= m_e^2 + m_\phi^2 -2E_1 E_\phi +2 \sqrt{E_1^2 -m_e^2} \sqrt{E_\phi^2 -m_\phi^2 } \cos\theta_\phi\,.\nonumber
\end{align}
In the derivation of the squared amplitude, one needs to define an additional Lorentz-invariant variable $p_{65} \equiv p_{\phi^*} \cdot p_\phi$ so that  every  scalar product of momenta can be written in terms of a combination of Lorentz-invariant variables~\cite{Chu:2018qrm}.
Using the fact that in 4-dimensional space-time, any five 4-vectors cannot be linearly independent, we express $p_{65}$ as a function of other Lorentz-invariant variables by solving ${\rm det}(M) = 0$, where the $(i,j)$ entry of the $5\times 5$ matrix $M$ is $p_i \cdot p_j$.
There are two solutions of $p_{65}$ corresponding to $\phi_{36}^* \in [0,\pi)$ and $\phi_{36}^* \in [\pi,2\pi)$.
Other Lorentz-invariant variables are not affected by $\phi_{36}^* \rightarrow 2\pi - \phi_{36}^*$.

\subsection{\boldmath$\phi$-$e$ scattering}
\label{Sec:scattering}

To leading order, the differential recoil cross section $d\sigma_{ \phi \text{-} e}/ dE_R$ as a function of the incoming $\phi$-energy, $E_\phi$, and the recoil energy of the electron, $E_R$, for the $F$ ($s$- and $u$-channel scattering) and $Z'$ ($t$-channel scattering) cases read as follows,
\begin{align}
	   \dfrac{d\sigma_{\phi \text{-} e}^{F}}{dE_R} &= \dfrac{c_L^2 c_R^2 (E_R +2 m_e)}{4\pi (E_\phi^2 - m_\phi^2)  m_F^2} \qquad (m_{e,\phi} \ll m_F), \\
	  \dfrac{d\sigma_{\phi \text{-} e}^{Z'}}{dE_R} &=  \dfrac{g_\phi^2 g_l^2 \big[2E_\phi m_e (E_\phi -E_R) - E_R m_\phi^2 \big] }{4\pi (E_\phi^2 -m_\phi^2)(2E_R m_e +m_{Z'}^2 )^2}\,,
\end{align}
where we use $g_L = g_R \equiv g_l$ for the sake of presenting a more compact formula.
The equations above are used in computing the electron scattering signal in mQ and BDX.

To properly account for the Pauli blocking factor in the computation of the upper boundaries of the SN1987A exclusion region, we also provide the differential cross sections in terms of Mandelstam variable $t$, taking again the limit of heavy mediators,
\begin{align}
	   \dfrac{d\sigma_{\phi \text{-} e}^{F}}{dt} &= \dfrac{c_L^2 c_R^2}{4 \pi  m_F^2 } \dfrac{4 m_e^2-t }{  (m_e^2+m_\phi^2-s)^2-4 m_e^2 m_\phi^2},\\
	   \dfrac{d\sigma_{\phi \text{-} e}^{Z'}}{dt} &= \dfrac{g_l^2 g_\phi^2 }{4 \pi m_{Z'}^4 }\,\dfrac{ (m_e^2+m_\phi^2-s)^2+t (s-m_e^2)}{ (m_e^2+m_\phi^2-s)^2-4 m_e^2 m_\phi^2}\,.
\end{align}

\section{Further 1-Loop diagrams}
\label{app:zwidth}

\begin{figure}[tb]
\vspace{.3cm}
\begin{center}
\includegraphics[width=0.5\columnwidth]{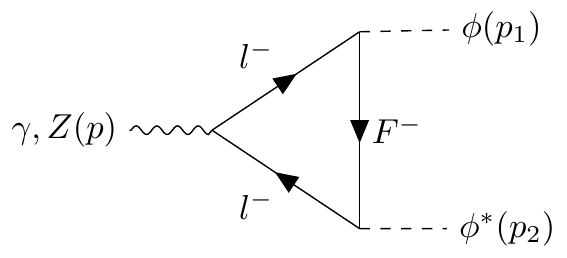}%
\includegraphics[width=0.5\columnwidth]{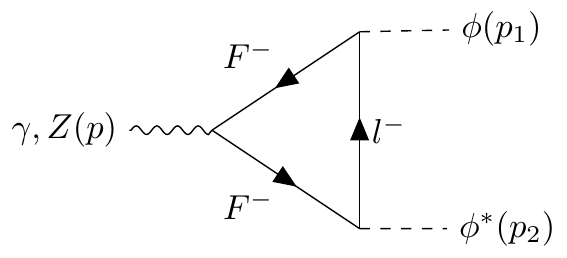}
\includegraphics[width=0.5\columnwidth]{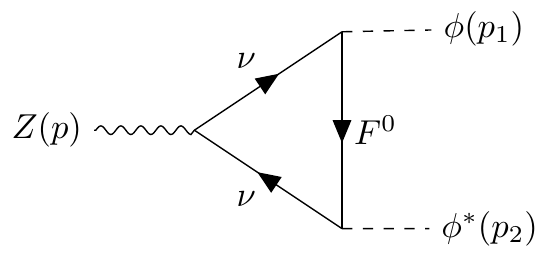}%
\includegraphics[width=0.5\columnwidth]{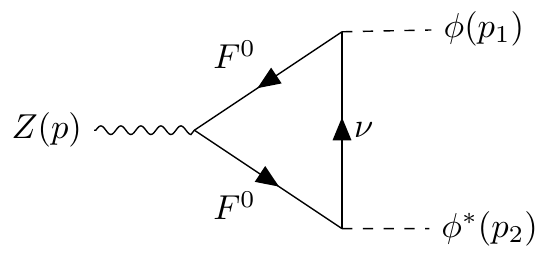}
 \includegraphics[width=0.45\columnwidth]{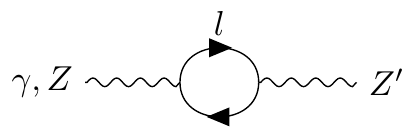}
\end{center}
\vspace{-.2cm}
\caption{%
{\it Top panel:} Triangle loop coupling $\phi$ to $Z$ or $\gamma$ with one charged $F$ (left) and two charged $F$ (right). {\it Middle panel:} Triangle loop coupling $\phi$ to $Z$ with one (left) or two (right) neutral $F$ states. {\it Bottom panel:} 
Loop-induced $\gamma$-$Z'$ and $Z$-$Z'$ mixing (which also contributes to the coupling to a $\phi$-pair.)}
\label{Fig:charged_Floop}
\end{figure}

The presence of heavy new charged fermions,  $F^\pm$, induces  effective interactions  between $\phi$ and SM gauge bosons through a set of triangle loops containing $F^\pm$ and SM charged leptons, demonstrated in Fig.~\ref{Fig:charged_Floop}.
This loop-induced charge radius interaction can e.g.~be probed in direct detection experiments~\cite{Bai:2014osa,Hamze:2014wca}. 

Here we detail the calculation of the coupling of the $\phi$-pair to an off-shell photon via the aforementioned triangle loops.  
The amplitude that needs to be dotted into the off-shell photon reads
\begin{equation}
i \mathcal{M}_{F,\mu} = \sum\limits_l \left( i\mathcal{M}^l_{1F, \mu} + i\mathcal{M}^l_{2F, \mu} \right)\,,
\end{equation}
where $i\mathcal{M}^l_{1F, \mu}$ and $i\mathcal{M}^l_{2F, \mu}$ correspond  to the diagrams containing one and two $F^\pm$ in the  loop, respectively. 
Note that we take one charged SM lepton as an example, and denote its mass by $m_l$. Here left- and right-handed SM leptons contribute equally, so we further set  $c_L=c_R \equiv c_F$. 

After using Feynman parametrization and dimensional regularization to perform the loop integral, we find that the divergences in $\mathcal{M}^l_{1F,\mu}$ and $\mathcal{M}^l_{2F,\mu}$ mutually cancel. The remaining finite terms read
\begin{widetext}
\begin{align}
   i \mathcal{M}^l_{1F,\mu}  &=  - i \dfrac{e c_F^2}{4 \pi^2 }  \int_0^1 dz \int _0^{1-z} dx  \, \left\lbrace p_{2, \mu}\left[ \dfrac{m_l^2 (z-1) +2 m_l m_F z + m_\phi^2 (z-1) z^2 + p^2 x(z+1)(x+z -1)}{ \Delta_{1F}}  +(3z+1) \ln \Delta_{1F} \right] \right. \notag \\
   & \quad\, \left. + p_{\mu} \left[  \dfrac{m_l^2 x +m_l m_F (2x-1) + m_\phi^2 z (xz-z-2x +1) + p^2 x(x-1)(x+z -1)}{ \Delta_{1F}}  +(3x-2) \ln \Delta_{1F} \right] \right\rbrace, \\
   i \mathcal{M}^l_{2F,\mu}  &=   i \dfrac{e c_F^2}{4 \pi^2 }  \int_0^1 dz \int _0^{1-z} dx  \, \left\lbrace p_{2, \mu}\left[ \dfrac{m_F^2 (z-1) +2 m_l m_F z + m_\phi^2 (z-1) z^2 + p^2 x(z+1)(x+z -1)}{ \Delta_{2F}}  +(3z+1) \ln \Delta_{2F} \right] \right.  \notag \\
   & \quad\, \left. + p_{\mu} \left[  \dfrac{m_F^2 x +m_l m_F (2x-1) + m_\phi^2 z (xz-z-2x +1) + p^2 x(x-1)(x+z -1) }{ \Delta_{2F}}  +(3x-2) \ln \Delta_{2F} \right] \right\rbrace,
\end{align}
\end{widetext}
with $p_\mu$ being the four-momentum of the photon, $p_{2,\mu}$ the four-momentum of one of the outgoing $\phi$ particles, and
{
\medmuskip=0mu
\thinmuskip=0mu
\thickmuskip=0mu
\nulldelimiterspace=1pt
\scriptspace=0pt
\begin{align*}
	& \Delta_{1F} = - m_l^2 (z-1) +z \big[m_F^2 + m_\phi^2 (z-1)\big] + p^2 x (x+z-1)\,, \nonumber \\
	& \Delta_{2F} = - m_F^2 (z-1) +z \big[m_l^2 + m_\phi^2 (z-1)\big] + p^2 x (x+z-1)\,.
\end{align*}
}
Note that if $m_l = m_F$, the total amplitude vanishes, as a manifestation of Furry's theorem. 
One can also directly write the combination of the two diagrams as an effective charge radius operator if we integrate out the heavy $F$,
\begin{equation}
 - b_\phi \partial_\mu \phi \partial_\nu \phi^* F^{\mu \nu}\,,
\end{equation}
where $F^{\mu \nu}$ is the field strength tensor of SM photon.
We have checked that our loop calculation correctly matches onto this effective operator, with the Wilson coefficient
\begin{widetext}
\begin{equation}
    b_\phi = \dfrac{e c_F^2}{4\pi^2} \int_0^1 dz\int_0^{1-z} dx\,\left\lbrace \dfrac{x(x+z-1)\big[m_l^2 (z^2 +z -2)+2 m_l m_F z - z(z+1)m_F^2 -z (z-1)m_\phi^2\big]}{\big[m_l^2 (z-1) -z m_F^2 -z (z-1) m_\phi^2\big]^2} - (m_l \leftrightarrow m_F) \right\rbrace,
\end{equation}
\end{widetext}
in numerical agreement  with previous results~\cite{Bai:2014osa,Hamze:2014wca}.  Note that a sum over the contributions from all SM leptons needs to be performed in the actual evaluation.

The above calculation can be generalized to infer the additional contribution to the invisible $Z$ decay width by replacing the relevant couplings in the above amplitudes by the weak charges 
\begin{equation}
\label{Eq:gammatoZ}
    e \,\rightarrow\, \dfrac{e}{2s_W c_W} g_V \,,
\end{equation}
where $s_W = \sin \theta_W$,  $c_W = \cos \theta_W$ with $\theta_W$ being the weak angle, and $g_V $ being the usual vector coupling of weak current; axial vector currents do not contribute when $c_L=c_R$ .
Note that one also needs to include the contribution from diagrams containing neutral leptons as shown in the bottom panel of Fig.~\ref{Fig:charged_Floop}. 
We have checked that the resulting bound on $c_F$ is rather weak, and is thus not shown in the constraint plots.

In the $Z'$ case, the SM photon or $Z$ boson  gains an effective coupling to $\phi$ via mixing with $Z'$, originating from a SM lepton loop, shown in the bottom panel of  Fig.~\ref{Fig:charged_Floop}.
To estimate the mixing, we need to compute the polarization mixing tensor $i\Pi^{\mu\nu}$ given by the usual Lorentz structure, $i\Pi^{\mu\nu} = i \left(p^2 g^{\mu\nu} - p^\mu p^\nu \right) \Pi (p^2)$.
For the SM photon and taking $g_L = g_R \equiv g_l$, we find in dimensional regularization the mixing self-energy
\begin{equation}
\label{Eq:selfenergy}
    \Pi (p^2) = - \dfrac{e g_l }{2 \pi^2} \int_0^1 dx\, x(1-x) \left[ \dfrac{2}{\epsilon}+ \ln \dfrac{\tilde{\mu}^2}{\Delta}+\mathcal{O}(\epsilon)\right],
\end{equation}
where $\epsilon$ is an infinitesimal number, $\tilde{\mu}^2 = 4\pi e^{-\gamma_E} \mu^2$ with $\gamma_E$ the Euler-Mascheroni constant, $\mu$ the renormalization scale, and $\Delta = m_l^2 -x(1-x) p^2$.

Equivalently, one can also re-write the self-energy of  Eq.~\eqref{Eq:selfenergy} in terms of the standard Passarino-Veltman integrals~\cite{Passarino:1978jh,tHooft:1978jhc,Hahn:2000jm} as 
\begin{equation}
    \Pi (p^2) = \dfrac{e g_l}{12\pi^2 p^2} \big[2A_0 (m_l^2) -(p^2 +2 m_l^2 ) B_0 (p^2, m_l^2, m_l^2)\big]\,.
\end{equation}
The divergence in $\Pi (p^2)$ can be cancelled by a counterterm, that is,  after specifying a renormalization condition, the effective mixing can be evaluated.
The mixing between $Z$ and $Z'$ is computed in the same way but with Eq.~\eqref{Eq:gammatoZ}.
Nevertheless, the ensuing constraint on $\sqrt{g_\phi g_l}$ from the $Z$ invisible decay is fully covered by the LEP bound.

\bibliography{refs}

\end{document}